\documentstyle[aps,prl,twocolumn,epsfig]{revtex}

\begin{document}
\draft

\twocolumn[\hsize\textwidth\columnwidth\hsize\csname@twocolumnfalse\endcsname%

\title{L\'{e}vy scaling: the Diffusion Entropy Analysis applied 
        to DNA sequences}
    \author{Nicola Scafetta$^{1,2}$, Vito Latora$^{3}$,
    and Paolo Grigolini$^{2,4,5}$.}
    \address{$^{1}$Pratt School EE Dept., Duke University,  P.O. Box 90291, Durham, North Carolina 27708 }
    \address{$^{1}$Center for Nonlinear Science, University of North Texas,
    P.O. Box 311427, Denton, Texas 76203-1427 }
    \address{$^{2}$ Dipartimento di Fisica e Astronomia, Universit\`a di
    Catania, and INFN, Corso Italia 57, 95129 Catania, Italy}
    \address{$^{3}$Dipartimento di Fisica dell'Universit\`a di Pisa and
    INFM, Piazza Torricelli 2, 56127 Pisa, Italy}
    \address{$^{4}$Istituto di Biofisica CNR, Area della Ricerca di Pisa,
    Via Alfieri 1, San Cataldo 56010 Ghezzano-Pisa, Italy}
    \date{\today}
    \maketitle

\begin{abstract}
We address the problem of the statistical analysis of a time series 
generated by complex dynamics with a new method: the Diffusion 
Entropy Analysis (DEA) (Fractals, {\bf 9}, 193 (2001)). This method 
is based on the evaluation of the Shannon entropy of the diffusion 
process generated by the time series imagined as a physical source of 
fluctuations, rather than on the measurement of the variance of this 
diffusion process, as done with the traditional methods. We compare 
the DEA to the traditional methods of scaling detection and we prove 
that the DEA is the only method that always yields the correct 
scaling value, if the scaling condition applies.  Furthermore,  
DEA detects the real scaling of a time series without 
requiring any form of de-trending. We show that the joint use of DEA 
and variance method allows to assess whether a time series is 
characterized by L\'{e}vy or Gauss statistics.  We apply the DEA to 
the study of DNA sequences, and we prove that their  large-time 
scales  are characterized by L\'{e}vy statistics, regardless of 
whether they are coding or non-coding sequences. 
We show that the DEA is a reliable technique and, 
at the same time, we use it to confirm  the validity of the 
dynamic approach to the DNA sequences, proposed in earlier work.
\end{abstract}
\pacs{03.65.Bz,03.67.-a,05.20.-y,05.30.-d}
\vspace{0.5cm}
%
]
%

\section{Introduction}
\label{introduction}
The recent progress in experimental techniques of molecular 
genetic shas made available a wealth of genome data
(see, for example, Ref. \cite{genbank}), and raised the interest for
the statistical analysis of DNA sequences. The pioneer
papers mainly focused on the controversial issue of whether
long-range correlations are a property shared by both coding and
non-coding sequences or are only present in non-coding
sequences \cite{stanley1,wli,voss,stanley2}.
The results of more recent papers \cite{mohanti,audit}
yield the convincing conclusion that the former condition applies.
However, some statistical aspects of the DNA sequences are
still obscure, and it is not yet known to what an extent the
dynamic approach to DNA sequences proposed by the
authors of Ref. \cite{barbi} is a reliable picture
for both coding and non-coding sequences.
The later work of Refs. \cite{paolophdthesis} and
\cite{buiattino} established a close connection between long-range
correlations and the emergence of non-Gaussian statistics, confirmed
by Mohanty and Narayana Rao \cite{mohanti}.
According to the dynamic approach of Refs. \cite{barbi,allegro}
this non-Gaussian statistics should be L\'{e}vy,
but this property has not yet been assessed with compelling evidence.
The reason for the confusion affecting this issue is deeper
than one can imagine, since it essentially depends
on the fact the there exists no reliable method of scaling
detection. In fact, all the traditional methods of
scaling detection on the market, the Detrended Fluctuation
Analysis (DFA) \cite{stanley3}, the Standard Deviation
Analysis (SDA) \cite{barbi},
and the Wavelets Spectral Analysis (WSA) \cite{audit,arneodo},
are based on the evaluation of the variance of the process, and
therefore yield a scaling that is the correct one
only if the process under study is Gaussian.

The main purposes of this paper are:

1) To clarify the meaning of scaling as a form of thermodynamic 
equilibrium that can be reached
after a long time transient, throughout which the conventional 
techniques of analysis can yield
misleading information.

2) To show that a new method, the Diffusion Entropy Analysis (DEA),
recently proposed in Ref. \cite{nicola}, is able to
yield the correct scaling, even when the observed
diffusion process is not Gaussian.
We shall show that the departure of the correct scaling, detected by means of the DEA, from the results of the  traditional methods,  all of them being variance-based
methods,  is a clear indication of the non-Gaussian character of the
process under study.

3) To show the DEA in action by means of an application  
to the study of DNA sequences.
As a remarkable result, we shall show that both coding and non-coding DNA
sequences depart from Gaussian statistics and
produce L\'{e}vy diffusion. This will shed light on some still obscure
aspects of the statistical properties of DNA.

\section{The Meaning of Scaling}
\label{sectscal}
The reason for the confusion still present in the issue of the
extraction of the long-range statistical properties
of DNA sequences (and more in general of any time series: heartbeats,
earthquakes, oscillations of markets stocks etc.)
is essentially due to the fact the there
are no reliable methods of scaling detection.
To clarify this crucial aspect we need to discuss, first, 
what scaling is all about. Scaling is a property of
a probability distribution $p(x,t)$, which formally reads as: 
\begin{equation}
\label{truscalff}
p(x,t) =\frac{1}{t^{\delta}} F\left(\frac{x}{t^{\delta}} \right).
\end{equation}
When we deal with a time series or a generic sequence we need 
first to construct the probability distribution $p(x,t)$. 
In order to do so we convert with some method, for instance the one used
in this paper, the single sequence into many distinct
trajectories. These trajectories start at time $t = 0$ from $x=0$,
and then spread over the $x$-axis, as a result of their, partial
or total, random nature. Thus, rather than observing a single
trajectory, we are naturally led to evaluate the probability
of observing it. In other words, we rest, with theoretical
  or computational arguments, on the probability of finding
  the variable $x$ in the interval $[x, x+ dx]$ at time $t$, denoted by us
as $P(x, dx, t)$. The probability density, $p(x,t)$, is defined
  by $P(x, dx, t) /dx$ .
The meaning of Eq.(\ref{truscalff}) is that the process
is {\it stationary}, in spite of the fact that the probability
density $p(x, t)$ broadens with time. To stress this aspect,
let us focus our attention on the probability densities
$p(x, t_{1})$ and $p(x, t_{2})$, at two distinct times
$t_{1}$ and $t_{2}$, with $t_{1} < t_{2}$. Let us squeeze
the abscissa scale of the later distribution, $p(x, t_{2})$,
by the factor $R \equiv( t_{1}/t_{2} )^{\delta} < 1$, and then enhance
the intensity of the resulting distribution density
by multipliying it by the factor $1/R > 1$. If the
property of Eq.(\ref{truscalff}) holds true, then
the resulting distribution density is identical to
the former, $p(x, t_{1})$. This is equivalent to
interpreting
the distribution density as a form of equilibrium
distribution. This property is deeply related to the
foundation itself of statistical
mechanics\cite{foundationofstatisticalmechanics}.
In fact, in the case
where the diffusion trajectory is the superposition
of many uncorrelated fluctuations, the resulting
diffusion process is predicted by the Central Limit
Theorem (CLT) to be a Gaussian probability distribution,
a special form of canonical equilibrium, and we can
  refer ourselves to the transient process necessary
  for the CLT to work as a kind of transition from
dynamics to thermodynamics. In this sense the scaling
property of Eq.(\ref{truscalff}) must be interpreted
as a form of thermodynamic equilibrium. Note that in
  the case of ordinary statistical mechanics, when the
CLT applies, we have that
$\delta = 1/2$ and $F(y)$ is a Gaussian function of $y$.

According to the new field of  Science of
Complexity \cite{yaneer,mandelbrot},
a complex process is expected to yield the property of Eq.(\ref{truscalff}) 
with $\delta \neq 1/2$ and (or)
$F(y)$ being a form different from the Gaussian one 
(we shall discuss an example of this non-Gaussian
form in later sections). Thus, this raises the question 
of whether a non canonical equilibrium condition 
can be generated by sequences reflecting complex dynamics. 
We should consider three different possibilites: 

1) Mandelbrot\cite{mandelbrot} proposes {\it Fractional Brownian 
Motion} (FBM) as a condition exceeding the limits of ordinary 
statistical mechanics. 
This corresponds to the scaling condition of Eq.(\ref{truscalff})
with $\delta \neq 1/2$ while $F(y)$ keeps its Gaussian form.

2) Another possible form of violation, naturally stemming from the
Generalized Central Limit Theorem (GCLT) \cite{gnedenkokolmogorov},
rests on Eq.(\ref{truscalff}) with $\delta > 1/2$ and 
{\it F(y) being a L\'{e}vy function}, with the asymptotic property
$lim_{y \rightarrow \infty} F(y) = const/y^{1+ 1/\delta}$.
This means the occurrence of a disconcerting condition, where
the second moment of the distribution is infinite.
It is obvious that in practice real time series cannot produce this
condition, and that the distribution moments of the observed
diffusion process are always finite,
being an imperfect realization of the diffusion process with infinite
moments.

3) Finally, we should consider also the stretched Gaussians 
emanating from subdiffusion\cite{subdiffusion}. 
Actually, this kind of process is not explicitly
examined in this paper. 
We expect that in this case the standard techniques of 
scaling dectection might do better than in case 2), 
since the stretched Gaussians are characterized by finite moments. 
Therefore, we shall focus our attention on both case 1), 
where the standard techniques are expected to yield exact results, 
and on case 2), where the standard techniques are expected to fail. 

As we shall show in this paper,
all techniques currently adopted to detect scaling are
explicitly or implicitly based on the measurement of the second
moment of the distribution $p(x,t)$. Thus, the scaling revealed
by the ordinary techniques of analysis might depart from the
genuine scaling of the process under observation, if this is
an imperfect realization of a diffusion process with infinite
moments. To stress this crucial aspect we adopt for the
scaling parameter $\delta$ the symbol $H$, according to a
notation proposed by Mandelbrot to honor Hurst\cite{hurstbook}
(see also Ref. \cite{feders}). Notice that a  widely adopted
method to express the condition of Eq.(\ref{truscalff}) is given by 
\begin{equation}
     x \propto t^{\delta}.
\label{scalingdefinition}
\end{equation}
This way of expressing the scaling condition is the source 
of misleading procedures. In fact it is usually assumed 
that it is equivalent to
\begin{equation}
     <x^{2}(t) >^{1/2} \equiv \int_{-\infty}^{+\infty} x^{2}~p(x,t)dx  \propto t^{\delta}.
\label{secondmomentscalingdelta}
\end{equation} 
We think that it is much more appropriate to use the following notation
\begin{equation}
     <x^{2}(t) >^{1/2}  \propto t^{H},
\label{secondmomentscalingh}
\end{equation} 
leaving open the possibility that $H \neq \delta$. 

In this paper we show that the {\it Diffusion
Entropy Analysis} (DEA) \cite{nicola} {\it is the only technique 
yielding the correct scaling $\delta$ when the observed
diffusion process departs from the FBM condition. }
In fact all the other techniques, including
the Detrended Fluctuation Analysis (DFA) \cite{stanley3},
the Standard Deviation Analysis (SDA)\cite{barbi}, and the Wavelets
Spectral Analysis (WSA)\cite{audit,arneodo}, yield a scaling that would
be correct only in the FBM case. This is so because, as we
shall see, these techniques rest on variance to evaluate scaling.
All these techniques, whose limitations are bypassed by
the DEA,  are in a sense different versions of the same method,
to which we shall refer to as the Variance Method (VM).
The departure of the correct scaling, revealed by the DEA,
from the results of the VM is consequently a proof of the
non-Gaussian character of the process under study.

\section{The Diffusion Entropy Analysis (DEA)}
The Diffusion Entropy Analysis (DEA)
is based upon the direct evaluation of the Shannon
entropy of the diffusion process. In the continuous-space and 
continuous-time representation for the probability density $p(x,t)$, 
the Shannon entropy \cite{shannon} of the diffusion process  reads
\begin{equation}
     S(t) = - \int_{-\infty}^{\infty} dx \, p(x,t) \ln [p(x,t)].
     \label{continuousshannonentropy}
     \end{equation}
    To show how the DEA works, let us assume that $p(x,t)$ fits the
scaling condition of Eq.(\ref{truscalff}).
     Let us plug Eq.(\ref{truscalff}) into
	Eq.(\ref{continuousshannonentropy}). After a simple algebra, we get:
	\begin{equation}
	    S(\tau) = A + \delta \tau ,
	    \label{keyrelation}
	    \end{equation}
	    where
	    \begin{equation}
		A \equiv -\int_{-\infty}^{\infty} dy \, F(y) \, \ln [F(y)]
		\label{ainthecontinuouscase}
		\end{equation}
		and
		\begin{equation}
	     \tau \equiv \ln (t) .
	     \label{logarithmictime}
	     \end{equation}
Eq.(\ref{keyrelation}) shows that if the diffusion process scales with the
parameter $\delta$, the resulting diffusion entropy becomes
a linear function of the logarithm of $t$, with a slope equal
to $\delta$. {\it This  makes the slope measurement equivalent
to the scaling detection, independently of the form of
$F(y)$ }.

In the case of ordinary Brownian diffusion, $\delta = 1/2$ and $F(y)$ has the
following Gaussian form
\begin{equation}
F(y) = \frac{exp\left(-\frac{y^{2}}{2 \sigma^{2}}\right)}{\sqrt{2\pi
\sigma^{2}}}.
\label{gaussianform}
\end{equation}
Thus Eq.(\ref{continuousshannonentropy}) becomes
\begin{equation}
S(t) =  \frac{1}{2} \left[1 + \ln(2 \pi \sigma^{2})\right] + 
\frac{1}{2} \ln(t).
\label{linearincrease}
\end{equation}
In this case, we have assumed the system to be already in the scaling
regime state. More in general, we shall have to
address the problem of the transition from the dynamic to the
thermodynamic (scaling) regime.

\section{  L\'{e}vy Walk}
\label{sectlevy}
The artificial sequences that we shall use in this paper to show the 
merits of DEA and the limits of VM rests on a dynamic approach 
adopted years
ago to derive L\'{e}vy statistics\cite{allegro,elena}.
The importance of this approach to L\'{e}vy statistics is due  the
fact that it makes  possible, in principle, to use the same 
perspective as that adopted in Ref. \cite{bianucci}. Bianucci
\emph{et al.}\cite{bianucci} discussed the case of a system of 
interest interacting with another system called \emph{booster} rather 
than \emph{thermal bath}, to emphasize
that no assumption on its thermodynamic nature  was done. The basic 
aspect of the research project of Ref. \cite{bianucci} was that 
statistical mechanics, in that case ordinary statistical mechanics, 
had to be derived from merely dynamic rather than thermodynamic 
arguments. The same approach can be applied to the derivation of 
L\'{e}vy statistics, with only one significant difference:
the phase space of the booster rather than being fully chaotic,
as in the case of ordinary statistical mechanics,
is weakly chaotic\cite{zaslavsky}.
The phase space consists of chaotic and regular regions,
and the booster trajectory tends to sojourn for a long time at
the border between chaotic and ordered regions.
The waiting time distribution is an inverse power law, and,
for simplicity, we assume it to be given by
\begin{equation}
\label{levywalkdis}
\psi(t)=(\mu-1)\frac{T^{\mu-1}}{(T+t)^\mu}.
\end{equation}
We make the assumption
\begin{equation}
\mu > 2,
\label{basiccoondition}
\end{equation}
which ensures the mean waiting time $\tau_{M}$ to get the finite value
\begin{equation}
\tau_{M} = \frac{T}{(\mu -2)}.
\label{meantime}
\end{equation}
It is evident from this formula that the parameter $T$, as well as 
the power index $\mu$, determine the time duration of the
sojourn of the trajectory at the border between chaotic and ordered 
regions. This inverse power law form, and the resulting stickiness,
are naturally generated by the self-similar nature of the
borders\cite{zaslavsky}. We call these crucial subsets of the phase 
space \emph{fractal borders}.

Now, let us assume that one of the variables of the phase space, 
called $\xi$, is the generator of the fluctuations that are collected 
by the diffusing variable $x$.
Since the fractal borders have a finite size, when the trajectory 
sticks to one fractal border, the variable $\xi$ gets a value that 
depends on the trajectory position.
Let us make also the assumption that there are only two fractal 
borders, and that their size compared to that of the whole phase space
is so small that the variable $\xi$ gets only two distinct values, 
denoted by us as $W$ and $-W$.  As an example of Hamiltonian model
generating velocity fluctuations we have in mind the kicked rotor
in the so called accelerating state\cite{mori,zumofen,laux}.
The booster trajectory moves erratically in the chaotic sea between 
the two fractal regions,  and after a given time sticks to one of the 
two fractal regions. After an extended time spent in this fractal 
region it goes back to the
chaotic sea, and after a short diffusion process, it either goes back 
to the earlier fractal region or it goes to the other one.
Due to the power law nature of the waiting time distribution of Eq.
(\ref{levywalkdis}), the sojourn in the chaotic sea can be ignored. 
As a result of this dynamic process we shall get a
sequence such as $ W, W,W, W, ...-W,-W, -W,-W, ......W,W,W,....$. In 
this paper we set $W = 1$. This is an example of the time series 
under discussion in this paper. For simplicity, rather than deriving 
it running a dynamic system, as the kicked rotor
in the accelerating state\cite{mori,zumofen,laux},
we can directly generate the random sequence $\{\tau_{i}, \xi_{i}\}$
in the following way:
first the numbers $\tau_{i}$ are randomly drawn from the
the distribution of Eq.(\ref{levywalkdis}); then the value of
$\xi_{i}$ is established by tossing a coin, and it is assumed
that the variable $\xi$ gets the specific value $\xi_{i}$
for the whole time interval $\tau_{i}$.

To understand the connection between this kind of sequence and
L\'{e}vy statistics, we have to use the fluctuation $\xi$ to generate
diffusion by means of the following equation of motion:
\begin{equation}\label{dynammodeflw}
\stackrel{{\bf \cdot} }{x}(t)=\xi (t)~.
\end{equation}
As remarked earlier, $\xi$ is a 
dichotomous variable, i.e. $\xi=\pm 1$, where $1$ is a unit
of length. The solution of (\ref{dynammodeflw}) is given by
\begin{equation}\label{soldiffdynalw}
x(t)=x(0)+\int\limits_{0}^{t}dt'~\xi (t')~,
\end{equation}
and our final goal is to evaluate $<x^2(t)>$.

As pointed out by Zaslavsky\cite{zaslavsky},
the condition $\mu > 2$, assumed throghout this paper
(see Eq.(\ref{basiccoondition})),
ensures the stationary condition, which allows us to properly define
$\Phi_{\xi}(t)$, the normalized correlation function of the 
fluctuation $\xi$. This important dynamic property, 
according to the renewal theory\cite{geisel}, 
is related to $\psi(t)$ by
\begin{equation}
\Phi_{\xi}(t) = \frac{1}{\tau_{M}} \int_{t}^{\infty} (t'-t) \psi(t') dt',
\label{renewaltheory}
\end{equation}
where $\tau_{M}$ denotes the mean waiting time.
Using for $\psi(t)$ the expression of Eq.(\ref{levywalkdis}) we obtain
\begin{equation}
\Phi_{\xi}(t) = \left(\frac{T}{t + T} \right)^{\mu -2}.
\label{explicitexpression}
\end{equation}
In this case $\tau_{M}$ is given by Eq.(\ref{meantime}).
Squaring the expression for $x(t)$ given by Eq.(\ref{soldiffdynalw})
  and by using the stationary  and  dichotomous nature of the fluctuation
$\xi(t)=\pm  1$ (the latter yielding  $<\xi^2>=1$),
  it is easy to prove that the mean square displacement $<x^2(t)>$ is given by
\begin{equation}\label{sdlw2f}
\frac{d}{dt}<x^2(t)>=2  \int\limits_{0}^{t}dt'~\Phi_{\xi} (t-t')~.
\end{equation}
Finally, by using Eq.(\ref{explicitexpression}) we get: 
\begin{equation}\label{hurslw2}
\lim _{t \to \infty }<x^2(t)>\propto t^{2H},
\end{equation}
with
\begin{equation}
\label{hursvaluedls2}
H=\frac{4-\mu}{2} ~~~~~~~ if ~\mu < 3,
\end{equation}
and
\begin{equation}
\label{ordinarystat}
H = \frac{1}{2} ~~~~~~~~~~~ if ~\mu > 3
\end{equation}
It is therefore evident that $\mu = 3$ is the border between ordinary 
and anomalous diffusion.
As pointed out in Section \ref{sectscal}, this result can be trusted only
in the Gaussian case.

Let us see why this way of evaluating scaling needs some caution.
Thank to the condition of Eq.(\ref{basiccoondition}),
we can define the  number $N =  [t/\tau_{M}]$, where $[y]$
denote the integer part of $y$.
In the case  $t >> \tau_{M}$ the number $N$ becomes
virtually identical to the number of random drawings of
the numbers $\tau_{i}$ and $\xi_{i}$.
This is equivalent to drawing the $N$ numbers
$\eta_{i} = \xi_{i} \tau_{i}$.

1) In the case where the condition $\mu > 3$ applies, this
distribution has a finite second moment. Thus, we
can use the Central Limit Theorem (CLT), which yields a
Gaussian diffusion, and consequently,  $H = 1/2$, which
correctly reflects the scaling in this case.

2) In the case $2<\mu <3$, the second moment of this distribution
is divergent, thereby preventing us from using the CLT.
However, in this case we use the Generalized Central Limit Theorem (GCLT) 
\cite{gnedenkokolmogorov}.
As shown in Ref.\cite{mario}, this random extraction of numbers yields a
diffusion process, described by the probability distribution
$p_{L}(x,t)$, whose Fourier transform, $\hat p_{L}(k,t)$, reads
\begin{equation}
\hat p_{L}(k,t) = exp( b|k|^{\mu -1}t)
\label{fouriertransformoflevy}
\end{equation}
with
\begin{equation}
b = W (TW)^{\mu-2} sin[\pi (\mu-2)/2] \Gamma(3-\mu) .
\label{mario}
\end{equation}
The subscript $L$ stands for L\'{e}vy. 
The numerical simulations support this theoretical expectation \cite{mario}.
Note that this dynamic approach to L\'{e}vy statistics coincides
with the L\'{e}vy walk\cite{geisel}. The difference between
  L\'{e}vy walk and L\'{e}vy flight is well known. In the case of 
L\'{e}vy flight the random walker makes instantaneously jumps of 
arbitrary intensity. In the case
of   L\'{e}vy walk, instead,
it takes the random walker a time proportional to $|\eta_{i}|$ to make
a jump with this intensity. In the case of L\'{e}vy flight, the 
random walker makes jumps of intensity $|\eta_{i}|$ at regular time 
intervals.

We note that the scaling of Eq.(\ref{truscalff})
derives naturally from the joint
use of the assumption $x \propto t^{\delta}$ and norm conservation.
It is straightforward to show that within the Fourier representation
the norm conservation yields $\hat p_{L} (0,t) = 1$. On the other hand,
moving from $|k|$ to $|\kappa| = |k| t^{1/(\mu -1)}$
we obtain the time independent Fourier transform
$\hat p_{ti}(\kappa) = exp(-b|\kappa|^{\mu -1}$),
which fits the normalization condition, and yields 
the scaling
\begin{equation}\label{levywalkdelta}
\delta=\frac{1}{\mu-1}~,
\end{equation}
which has to be compared to Eq.(\ref{hursvaluedls2}). It is evident 
that $H \neq \delta$, in this case.

In this paper, we shall focus our attention on the dynamic
condition fitting both the condition of Eq.(\ref{basiccoondition})
$\mu >2$, and the condition
\begin{equation}
\mu < 3.
\label{divergentsecondmoment}
\end{equation}
This is in line with the arguments of the dynamic approach to DNA of 
the earlier work of 
Refs.\cite{barbi,paolophdthesis,buiattino,allegro}, which proved the
DNA sequences to be equivalent to a dynamic process fitting both 
conditions, ensuring \emph{stationarity}, the former, and 
\emph{superdiffusion}, the latter, at the same time.

There are two important issues to clarify before proceeding with the next
sections.
The reader can find a detailed account somewhere 
else\cite{allegro,gianmarco,mauro}. However, to make this paper as 
much selfcontained as possible,  we shall shortly outline both of 
them. The first issue has to do with the time required for the GCLT 
to apply.
The work of Ref. \cite{gianmarco} shows that the predictions of the 
GCLT are realized by the following expression for $p(x,t):$
\begin{equation} p(x,t) = K(t) p_{T}(x,t)
\theta (Wt-|x|) +  \frac{1}{2}\delta (|x|-Wt) I_{p}(t).
\label{truncated}
\end{equation}
where $p_{T}(x,t)$ is a distribution that for $t \rightarrow \infty$
becomes identical to the L\'{e}vy probability distribution of the 
variable $x$, namely a function whose Fourier transform coincides 
with Eq.(\ref{fouriertransformoflevy}), 
$\theta$ denotes the Heaviside step function and $K(t)$ is 
a time-dependent factor ensuring the normalization of the 
distribution $p(x,t)$. 
This contribution to Eq.(\ref{truncated}) 
is a truncated L\'{e}vy distribution, the rationale for it being that no trajectory can travel with velocity of intensity larger than  $W$. The trajectories that at time $t> 0$ are still travelling in the same direction as at time $t = 0$ produce two peaks located at the propagation fronts, $x = Wt$ and $x = -Wt$, and their contribution to $p(x,t)$ is given by the second term on the right hand side of Eq.(\ref{truncated}). The number of trajectories that contribute to the peaks is given by the function $I_{p}$ that has been evaluated in detail by the authors of Ref.\cite{gianmarco}. Here it is enough to say that these authors find 
\begin{equation}
\label{correlationfunction}
\lim_{t \rightarrow \infty} [I_{p}(t) - \Phi_{\xi}(t) ] = 0 .
\end{equation}
This means that in the time asymptotic limit the peak intensity 
becomes identical to the correlation function 
$\Phi_{\xi}(t)$ of Eq.(\ref{explicitexpression}). 
On the basis of these arguments they reach the conclusion that in the
asymptotic time limit Eq.(\ref{truncated}) becomes identical to
\begin{equation} p(x,t) =  p_{L}(x,t)
\theta (Wt-|x|) +  \frac{1}{2}\delta (|x|-Wt) \Phi_{\xi}(t),
\label{truncated2}
\end{equation}
which coincides with the earlier prediction of Ref. \cite{allegro}.
This conclusion seems to be compatible with the results obtained by using the theory of Continuous Time Random Walk (CTRW)\cite{ctrw}, although these authors do no refer explicitly to the correlation function $\Phi_{\xi}(t)$. For an earlier work based on the CTRW see Ref. \cite{chemphysklafter}

To provide an answer to the first question it is enough to rest
on the earlier result of Eq.(\ref{truncated2}).  It takes 
an infinite time for the GCLT to apply: in fact the intensity of the 
peaks of the propagation front is proportional to the correlation 
function of Eq.(\ref{correlationfunction}), which is not integrable. 
During this long transient, as we shall see, the DEA gets closer and 
closer to the true scaling of Eq.(\ref{levywalkdelta}), while the 
distribution second moment, which is finite due to the truncation of 
the L\'{e}vy distribution, yields the fake scaling of Eq.(\ref{hursvaluedls2}).

The second issue is less relevant to the main purpose of this paper.
It has to do with another approach to the true scaling, already
discussed in Ref. \cite{allegro}.
This has to do with the Hamiltonian derivation of L\'{e}vy statistics mentioned
in Section \ref{sectscal}.
We study  the time evolution of the probability distribution of
the diffusion variable $x$, of the fluctuating variable $\xi$
and of all other variables that might be responsible for the fluctuations
of $\xi$.
Then, we make a trace over all the ``irrelevant" variables,
namely, all the variables but $x$. The resulting equation of
motion is not Markovian, and no ordinary method to make the
Markovian approximation can be applied. This is so because
the projection method yields a time convoluted diffusion
equation with a memory term given by the correlation function
$\Phi_{\xi}(t)$ of Eq.(\ref{explicitexpression}), which
is not integrable. Consequently a new way to make the Markovian
approximation also in this case was invented\cite{allegro}. It was 
noticed that this approximation changes the time convoluted diffusion 
equation into a master equation\cite{juri}. To derive from it a 
result consistent with that of the CTRW used in an earlier work of 
Zumofen and Klafter \cite{ZuKla}, and with L\'{e}vy statistics as well,  
the authors of Ref. \cite{juri} had to use as a bridge the 
master equation method of Ref. \cite{katjia}.
This master equation gets 
the form of a fractional derivative, and, the resulting diffusion 
process
coincides with the predictions of the GCLT, with a diffusion strength 
$b$ that coincides with that of Eq.(\ref{mario}). It comes to be a 
surprise, therefore, that the recent work of Ref.\cite{mauro} proves 
that the exact solution of the time convoluted diffusion equation
yields the same scaling as the VM, namely, the scaling of Eq.
(\ref{hursvaluedls2}). This suggests that densities and trajectories 
might not speak the same language in the case of non-ordinary 
statistical mechanics,
and it makes much stronger than ever the need for  detecting the 
correct scaling of a time series.

\section{The algorithm}
\label{sectalg}
Let us consider a sequence of $M$ numbers
\begin{equation}
     \xi_{i} ,    \quad i = 1,  \ldots , M.
     \label{thesequenceundestudy}
     \end{equation}
     The purpose of the DEA  is to establish the possible
     existence of a scaling, either normal or anomalous, in the most
     efficient way as possible without altering the data with any form
     of detrending. Here we describe the algorithm adopted in this 
paper. 

Let us select first of all an integer number
$l$, fitting the condition $1 \leq l \leq M$ .
This integer number will be referred to by us as ``time''.
For any given time $l$ we can find $M - l +1$ sub-sequences
of length $l$ defined by
\begin{equation}
	    \xi_{i}^{(s)} \equiv \xi_{i + s}, ~~~  i = 1, \ldots , l   ,
	    \label{multiplicationofsequence}
\end{equation}
with  $ s = 0,  \ldots ,  M-l$.
For any of these sub-sequences we build up a diffusion trajectory,
$s$, defined by the position
\begin{equation}
	x^{(s)}(l) = \sum_{i = 1}^{l} \xi_{i}^{(s)}
	= \sum_{i = 1}^{l} \xi_{i+s}.
	    \label{positions}
\end{equation}

	    Let us imagine this position as that of a Brownian particle that
	   at regular intervals of time has been jumping forward of
	   backward according to the prescription of the corresponding
	   sub-sequence of Eq.(\ref{multiplicationofsequence}). This means
	   that the particle before reaching the position that it holds at
	   time $l$ has been making $l$ jumps. The jump made at the
	   $i$-th step has the intensity $|\xi_{i}^{(s)}|$ and is forward or
	   backward according to whether the number $\xi_{i}^{(s)}$ is
	   positive or negative.

	   We are now ready to evaluate the entropy of this diffusion process.
	   In order to do so we have to partition the $x$-axis 
           into cells of size $\epsilon(l)$ and to count how many 
           particles are found in the cell $i$ at a
	   given time $l$. We denote this number by $N_{i}(l)$. Then
	   we use this number to determine the probability that a particle
	   can be found in the $i$-th cell at time $l$, $p_{i}(l)$, by means
	   of
	   \begin{equation}
	    p_{i}(l) \equiv  \frac{N_{i}(l)}{(M-l+1)} .
	    \label{probability}
	    \end{equation}
	    The entropy of the diffusion process at time $l$
	    is: 
\begin{equation}
  S_{d}(l) = - \sum_{i} p_{i}(l) \ln [p_{i}(l)] .
\label{entropy}
\end{equation}
Note that the subscript $d$ stands for \emph{discrete} and serves the 
purpose of
reminding the reader that the numerical evaluation of the diffusion 
entropy departs by necessity from the continuous-time and
continuous-space picture of Eq.(\ref{continuousshannonentropy})
The easiest way to proceed with the
choice of the cell size, $\epsilon(l)$, is to assume it independent
of $l$ and determined by a suitable fraction of the square root of
the variance of the fluctuation $\xi_i$.

In this paper we study sequences of numbers $\xi_i = +1$ or $-1$. 
Because at any step, the jump has the intensity equal to 1, the most 
reasonable choice of the cell size is given by $\epsilon(l)=1$. In 
this way any cell
corresponds to a unique position $x(l)$ of the diffusion trajectory
defined in (\ref{multiplicationofsequence}) and (\ref{positions}).
Moreover, $\epsilon(l)=1$ is the square root of the variance of
the random dichotomous fluctuation $\xi_i$ of intensity equal to 1.

Few remarks about the meaning of the integer number $l$
are necessary for the reader to understand the
content of the next sections.
As said before, $l$ is the length of a window moving
all over the available sequence to define distinct trajectories.
These trajectories are used to produce diffusion, and
consequently we shall often refer to $l$ as time. This should not 
confuse the reader. The adoption of the term time is suggested by the 
formal equivalence with the processes of either normal or anomalous 
diffusion, where walker's jumps occur in time. Here, these jumps 
occur as we move from a sequence site to the next, and consequently 
time here has to do with the site positions. Furthermore, we shall be 
often using for this kind of time the symbol $t$ rather than $l$. 
This has to do with the fact that for windows of very large size the 
integer number $l$ becomes virtually indistinguishable from a 
continuous number. To emphasize this aspect we shall adopt the symbol 
$t$ rather than $l$.

\section{Transition regime:  Random Walk and L\'{e}vy walk}
In Section \ref{sectscal} we have shown that scaling is equivalent
to thermodynamic equilibrium with the equilibrium distribution $F(y)$.
We refer to the transient process necessary to realize this form of 
thermodynamic equilibrium from the initial condition with all the 
trajectories located at $x = 0$, as transition
from microscopic dynamics to thermodynamics.  Here we illustrate this 
transition in two different cases,  ordinary Brownian motion and 
L\'{e}vy walk. In the former case the transition from microscopic 
dynamics to thermodynamics can be interpreted as a transition from 
the discrete to the continuous time representation.
In the second case the transition is more extended and can be still 
perceived after reaching the continuous time regime.

\subsection{The transition regime in the case of Brownian walk}
The discrete perspective can be illustrated by using the random
walk theory that is expected to apply when our dichotomous
signal is completely random.
In this specific case, with no correlation, the probability $p_m(l)$,
   for the random walker to be at position $m$ after $l$ jumps of intensity $1$
   in either positive or negative direction, is determined by the
   binomial expression \cite{Reichl}:
\begin{equation}\label{prorandwalker}
p_m(l)=\frac{1}{2^l}{l \choose \frac{l+m}{2}}\frac{1+(-1)^{l+m}}{2}.
\end{equation}
and the diffusion entropy  reads
	 \begin{equation}
	S_{d}(l) = -  \sum_{m = -l}^{l} p_m(l) \, ln[p_m(l)].
	     \label{exactinthediscrete}
	     \end{equation}
In the continuous time limit we expect Eq.(\ref{linearincrease}) to apply. 
Fig.1 shows that, after a short initial regime, the discrete diffusion entropy 
converges to the continuous time prescription (solid line in figure). 
In the case of Brownian walk we can interpret the transition from 
microscopic dynamics to thermodynamics as the transition from
the binomial formula of Eq.(\ref{prorandwalker})  to the Gaussian expression
of Eq.(\ref{gaussianform}), with $\sigma = 0.5$.

\subsection{The transition regime in the case of L\'{e}vy walk}
\label{secttrans}
Here we show how to build a sequence corresponding to the prescription of
Section \ref{sectlevy}.
In an earlier work\cite{luigi} the reader can find the illustration 
of an algorithm that, using a generator of random numbers of the 
interval $[0,1]$, creates the waiting time distribution of 
Eq.(\ref{levywalkdis}). Here we illustrate a slightly different 
method,  generating a distribution of integer times that is exactly, 
rather than approximately, equivalent to a shifted inverse power law. 
This serves the purpose of making as fast as possible the transition 
from microscopic dynamics to thermodynamics, without further delay 
caused by the time it takes the distribution to become the shifted 
inverse power law of Eq.(\ref{levywalkdis}).

To realize this purpose, first of all we need to generate
a series of $i$ integer numbers $L(i)$ according to a probability
distribution $p(L)$: these numbers can be interpreted as the lengths of
strings of the sequence to build up.
Then, for any string, we toss a coin and we  fill it entirely with
$+1$'s or $-1$'s, according to whether we get head or tail.
We assign to the integer numbers L(i) the following inverse power law:
\begin{equation}
\label{powerlaw}
p(L)=\frac{C}{(T+L)^{\mu}},
\end{equation}
where $T$ and $C=\left(\sum _{L=1}^{\infty } \frac{1}{(T+L)^{\mu}}\right)^{-1}$
are two constants related the one to the other in such a way as to 
realize the normalization condition without  the continuous time 
assumption behind Eq.(\ref{levywalkdis}).
It is evident that in the asymptotic limit of very large times the 
distribution of Eq.(\ref{powerlaw}) becomes equivalent to that of Eq.
(\ref{levywalkdis}).

To create the distribution of Eq.(\ref{powerlaw})
we proceed as follows.
We divide the interval  of  real numbers  [0,1] in infinite sectors.
The L-{\it th} sector, $R_L$, covers the space
\begin{equation}\label{sectorsp}
R_L \equiv \left[ X(L), X(L) +\frac{C}{(T+L)^{\mu}} \right),
\end{equation}
where
\begin{equation}
X(L) =  \left\{
\begin {array} {ccl}
0 & { \rm if } & L=1, \\
C \sum _{n=1}^{L-1} 1/(T+n)^{\mu}& { \rm if } & L  > 1.
\end{array}
  \right.
\label{positionsX}
\end{equation}
The length of the  sector $R_L$ is equal to the probability
$p(L)$ given by the Eq.(\ref{powerlaw}).
Then, by using a computer, we generate a sequence of
rational random numbers $\Upsilon(i) $ uniformly distributed
between 0 and 1:
if the rational number $\Upsilon (i)$ belongs to
the  sector $R_L$, the value $L$ will be assigned to
the element $L(i)$ of the  sequence of integer numbers.
The described algorithm and the uniformity of the sequence
of rational random numbers $\Upsilon(i) $ 
assure that the sequence of integer numbers $L(i)$ is distributed
exactly according to the power law given by the equation (\ref{powerlaw}).
It is worth to point out that this special method of creating the 
artificial sequence to analyze by means of the DEA is equivalent to 
that used  by Zumofen and Klafter \cite{ZuKla}.
Of course, due to the time asymptotic equivalence with the condition discussed
in Section \ref{sectscal},
even in this case the thermodynamic regime is characterized by 
L\'{e}vy statistics
and the proper scaling is that of Eq.(\ref{levywalkdelta}). The 
diffusion entropy, $S_d(l)$
of Eq.(\ref{entropy}), is expected  to converge asymptotically to the 
curve of Eq.(\ref{keyrelation}).
For example, if we set $\mu = 2/3$ in Eq.(\ref{powerlaw}),
on the basis of Eq.(\ref{levywalkdelta}) we expect a value $\delta = 2/3$.
In principle, using the theoretical approach of Zumofen and 
Klafter\cite{ZuKla} we might also evaluate the value of $A$ using
Eq.(\ref{ainthecontinuouscase}). Since this is not very relevant for 
the present paper, 
we skip this issue, and we rest on the numerical 
simulation to conclude that $A = 1$, thereby reaching the conclusion
that the asymptotic time limit is well reproduced by
\begin{equation}\label{Levyfitting}
S_{d}(t)=1+\frac{2}{3}  \ln(t)
\end{equation}
 From Fig. 2 we see indeed that Eq.(\ref{Levyfitting}) fits remarkably 
well the time asymptotic limit of the numerical curve.
We see that this limiting condition is reached after a transient that 
is significantly larger than that of Fig. 1. A satisfactory 
discussion of this transient will be presented
in Ref.\cite{gianmarco}.

\subsection{DEA and SDA at work}
\label{subsda}
In this Subsection we show the benefit of the joint use of DEA and 
SDA. The standard deviation at the diffusion time $l$, $D(l)$, rests on the following 
prescription
\begin{equation}
\label{standarddeviation}
D(l) = \sqrt{\frac{\sum_{n = 1}^{M - l}(x_{n}(l) - \bar x)^{2}}{M - l-1}},
\end{equation}
where, according to the notation of Section \ref{sectalg}, $M$ is the 
sequence length, $ l $ denotes the width of moving windows necessary
to create distinct trajectories and $\bar x$ denotes the mean value of $x(l)$.

According to the theoretical remarks of Section V, the adoption 
of this method applied to an artificial sequence
generated by the inverse power law distribution 
of Eq.(\ref{powerlaw}), with $2 \leq \mu \leq 3$ should yield 
the true scaling of Eq.(\ref{levywalkdelta}). 
The SDA should generate the Hurst scaling of Eq.(\ref{hursvaluedls2}). 
We make the analysis of five artificial sequences
with the power indices: $\mu$ = 2.8, 2.6, 2.5, 2.4, 2.2. 
We note that at both $\mu = 3$ and $\mu =2$ the two predictions 
yield the same values $\delta = H = 0.5$ and $\delta = H = 1$, respectively. 
Therefore we focus our attention on the intermediate values of $\mu$. 
For these intermediate values
the correct scaling, namely the L\'{e}vy scaling, yields 
$\delta$ = 0.556, 0.625, 0.667, 0.714, 0.833, 
respectively, while the Hurst scaling is expected to be
$H$ = 0.6, 0.7, 0.75, 0.8, 0.9, respectively. For the sake of 
reader's convenience this situation is summarized in Table I. 
The numerical results illustrated in Fig. 3 provide a strong 
support to the theoretical arguments of Section V, 
and to our claim about the accuracy of the DEA.
In fact, we see that the DEA yields a remarkable 
agreement with the L\'{e}vy scaling, while the scaling 
detected by the SDA virtually coincides with the Hurst scaling.

In general, when the secrete recipe driving the sequence under study 
is not known, the comparison between the DEA and SDA results plays 
an important role
to assess the statistical nature of the process. In fact, in the 
case of L\'{e}vy statistics, 
it is easy to show, using  Eq.(\ref{hursvaluedls2}),
that $\delta$ is related to $H$ by
\begin{equation}
         \delta = \frac{1}{(3 - 2 H)}    .
     \label{LCcond}
\end{equation}
In the FBM   case, according to theoretical arguments of
Section \ref{sectscal}, we have
\begin{equation}
        \delta = H   ,
     \label{GCcond}
\end{equation}
and this equality can be considered as a plausible indication that 
the Gausssian condition applies.
The results of Fig.3 fit Eq.(\ref{LCcond}), 
thereby confirming the L\'{e}vy nature of the diffusion process.

\section{Applications to DNA sequences}
In the last few years, thanks to the recent progress in
experimental techniques in molecular genetics, a wealth of genome
data has become available (see for example Ref.\cite{genbank}).
This has triggered a large interest both in the study of mechanics
of folding \cite{torcini}, and on the statistical properties of
DNA sequences. In particular, genomes can be considered as long
messages written in a four-letter alphabet, in which we have to
search for information (signal).
Recently, there have been many papers pointing out that DNA
sequences are characterized by long-range correlation, this being
more clearly displayed by non-coding than by coding sequences
\cite{stanley1,stanley2,allegro,stanley3}.

In this section we will study a large sample of DNA sequences (a dozen 
of both coding and non-coding sequences). 
In particular we discuss in detail three DNA sequances: 
\\
-- the human
T-cell receptor alpha/delta locus (Gen Bank name HUMTCRADCV)
\cite{stanley3}, a {\it non-coding} chromosomal fragment (it
contains less than 10\% coding regions); 
\\
-- the Escherichia Coli K12
(Gen Bank name ECO110K) \cite{stanley3}, and the Escherichia Coli
(Gen Bank ECOTSF) \cite{buiattino}, two genomic fragments
containing mostly {\it coding regions} (more than 80\% for
ECO110K). 
\\
The three sequences have  comparable lengths,
  $M=97634$   basis for  HUMTCRADCV, $M=111401$  basis
for  ECO110K and $M=91430$   basis for  ECOTSF, respectively.
The first two sequences have been analyzed in Ref. \cite{stanley3}
by means of the  Detrended Fluctuation (DFA). The fundamental
difference between them is that the non-coding sequence, namely
HUMTCRADCV, shows the presence of long-range correlation at all
scales, while the sequence ECO110K, a coding sequence, shows the
presence of long-range correlation only at the short-time scale.
The third  sequence, ECOTS, has been studied in Ref.
\cite{buiattino} with the interesting conclusion that the
large-time scale shows  non-Gaussian statistics. The authors
of Ref. \cite{stanley3},  using the illuminating example of the
lambda phage genome, pointed out that the DFA does not mistake
the presence of patches of different strand bias for
correlation. This
is an important property, shared by the DEA  method, which is widely
independent of the presence of biases, since the entropy increases
mainly as a consequence of the trajectories departing from one another.
In this Section we show that 
{\it the DEA method makes it
possible to relate the non-Gaussian statistics and the anomalous
scaling of the large-time scale to the same cause: the onset of
L\'{e}vy statistics }.

\subsection{The numerical representation of DNA}
The usual way to study the statistical properties of DNA is to
consider a sequence of four bases: adenine, cytosine, guanine, and
thymine (respectively A, C, G, and T), at the simplified level of
a dichotomous sequence of two symbols, purine (for A and G) and
pyrimidine (for C and T). A trajectory, the so-called DNA walk,
can be extracted by considering a one-dimensional walker
associated to the nucleotide sequence in the following way: the
walker takes one step up when there is a pyrimidine in the
nucleotide and a step down when there is a purine. The DNA
sequence is therefore transformed in a sequence $\xi_{i},~~i =
1,...,M$, of numbers $+1$ or $-1$.

As pointed out at the end of Section
\ref{sectalg}, we associate the site position along the sequence
with time.
Thus, $i$ is conceived as a discrete time,
and the walker makes a step ahead or
backward, according to whether at time $i$ the random walker
sees $+1$ or $-1$, namely if the i-th site of the DNA sequence
hosts a pyrimidine or a purine. The displacement of the walker
after $l$ steps is $x(l)= \sum_{i=1}^l ~ \xi_i$ and is reported in
Fig. 4 for the three sequences under consideration.

\subsection{The three variance methods (VM) at work: 
non-coding and coding DNA sequences}
This section is devoted to illustrating the three different
realizations of the variance method (VM),
namely,
the Detrended Fluctuation Analysis (DFA) \cite{stanley3},
the Standard Deviation Analysis (SDA) \cite{barbi},
and the Wavelets Spectral Analysis (WSA) \cite{audit,arneodo}.
We have already discussed the first two methods in the previous sections. 
We have also showed some results of the application of SDA to an 
artificial sequence
in section \ref{subsda}.
As to the WSA, it was first adopted to study DNA sequences by
Arneodo and collaborators in Ref. \cite{arneodo}, and it consists in
reporting the square root of the wavelet variance.
In this way, the scaling is comparable to those detected by
DFA and SDA, and, as we shall see, it gives  indeed the same results.

The first property  we notice is that all the three series present
``patches'', i.e. excess of one type of nucleotide. In the DFA 
of ref. \cite{stanley3}, Stanley and collaborators
adopt a detrending procedure to detect the true scaling,
since the steady bias hidden in the data can produce effects
which might be mistaken for a striking departure
from Brownian diffusion, while the interesting form of scalings
must be of totally statistical nature. They
define a detrended walk by subtracting the local trend from
the original DNA walk and then they study the variances
$F(l)$ of the detrended walk.
If the walk is totally random, as in the ordinary Brownian
motion, no correlations exist and $F(l) \sim l^{1/2}$. On the
contrary, the detection of $F(l) \sim l^{H}$ with either $H>1/2$
or $H < 1/2$ is expected to imply the presence of extended
correlation, which, in turn, is interpreted as a signature of the
complex nature of the observed process.

To illustrate the results of these authors, let us limit to the 
long-time region the adoption of the symbol
$H$, which, according to Section II, is used by us to denote the 
scaling emerging from the VM.
When the VM method is applied to the short-time region let us call 
the scaling parameter determined by the VM with the symbol $H'$.
Stanley \emph{et al.}\cite{stanley3} found a
scaling exponent $H'=0.61$ for the non-coding intron sequence
HUMTCRADCV, and $H'=0.51$ for the intronless sequence ECO110K.
They claim that their detrending method is able
to avoid the spurious detection of apparent long-range
correlations which are the  artifacts of the patchiness.

We are now ready to show the three methods at work on the
DNA data sets we want to study in this paper.

{\bf Non-coding DNA}. 
Fig.5 refers to the sequence HUMTCRADAVC and shows that,
within the statistical error,
the three VM techniques yield the same long-time scaling,
more precisely the three scaling exponents $H$ obtained are
$0.59\pm0.01$ (SDA), $0.60\pm0.01$ (DFA), $0.61\pm0.01$ (WSA).

{\bf Coding DNA}. 
In Figs. 6 we study the two sequences ECO110K and ECOSTS, and
we show that the same equivalence applies to both short-time
and long-time scaling.
In fact for both sequences we find that
$H'$ is $0.53\pm0.01$ (SDA), $0.52\pm0.01$ (DFA),
$0.52\pm0.01$ (WSA), and that $H$ is
$0.73\pm0.01$ (SDA), $0.75\pm0.01$ (DFA), $0.74\pm0.01$ (WSA).

Before moving to illustrate the results obtained by
the DEA, some comments are in order. DFA detects the scaling in the long-time
region later because of the detrending that cuts off long local
trend. In Ref. \cite{stanley3}, Stanley and collaborators were
interested in studying the  scaling in the short-time region in order to
distinguish the non-coding from the coding DNA sequences. The DFA
aims at making more visible this regime. However, we think that it is 
more convenient to study the signal as it is, since detrending might
erase important information as well as the deceiving indication of a 
correlation that does not exists.

Figs. 5 and 6 show that there is no
difference between SDA and WSA. This is because the Wavelet
Transform behaves  like the Fourier Transform that
studies the variance of the signal. Therefore, WSA, as Fourier
Spectral Analysis, can detect the true scaling only in the
Gaussian case. In all other cases, WSA detects only the
variance scaling, and this, as pointed out in
Section \ref{sectlevy},  may not coincide with the true scaling.

\subsection{The Copying Mistake Map: a model for DNA sequences}
According to the dynamical model of
Ref. \cite{barbi} a {\it non-coding} DNA sequence corresponds to an
artificial sequence with  inverse power law long-range
correlation as the L\'{e}vy walk of Section \ref{sectlevy},
examined by means of the DEA in Section VI.
On the other side, a {\it coding sequence} can
be obtained by adopting a kind of generalization of the L\'{e}vy 
walk. This generalization becomes a model called Copying Mistake Map
(CMM) \cite{barbi}.
This model rests on two sequences of $+$'s and $-$'s, running 
independently the one from the other.
The former sequence is the correlated sequence studied in Section 
VI C by means of the joint use of DEA and SDA. 
The latter sequence is obtained by tossing a coin. 
According to the  CMM, the generic $i-th$ site of the DNA sequence 
is assigned the symbol pertaining to the $i-th $ site of either 
the former or the latter sequence. 
The former sequence is selected with probability $p_{L}$ 
and the latter sequence with probability $p_{R} = 1 - p_{L}$.  
In the case of coding sequences usually the condition
\begin{equation}
    p_{R} \gg p_{L}
    \label{xxx}
\end{equation}
applies.
The authors of Ref. \cite{buiattino} pointed out that the
CMM model is equivalent to an earlier model \cite{stanley2,araujio}
called Generalized L\'{e}vy Walk (GLW).
The CMM (and the GLW, as well, of course) yields, for short
times, a diffusion process indistinguishable from ordinary Brownian
motion. At large times, however, the long-range correlation
predominates. In Ref. \cite{buiattino} the CMM was adopted
to account for the properties of prokaryotes,
for which a significant
departure from Gaussian statistics occurs.
One of the coding sequences studied here, namely ECOTSF,
is the same as one discussed in Ref. \cite{buiattino}.
It produces strong deviations from Gaussian statistics.
On the basis of that, and of the results of Section \ref{secttrans},
we expect also for coding sequences at large times a
scaling parameter $\delta$ corresponding to the L\'{e}vy statistics, 
and so, to the prediction of 
both Eq.(\ref{levywalkdelta}) and Eq.(\ref{LCcond}).

The CMM is a model flexible enough as to move from the Gaussian to 
the L{\'{e}vy condition. This is done simply setting $p_{R} = 0$. On 
the other hand, if the condition of Eq.(\ref{xxx}) applies, in the 
long-time limit we expect the condition of L\'{e}vy statistics will 
emerge again. This is so because the most evident sign of L\'{e}vy 
statistics is given by the power law character of the distribution 
tails. The correlated component of the CMM model results in a process 
of diffusion faster than ordinary diffusion, and so faster than the 
diffusion generated by the random component. As a consequence, the 
distribution tails are forced to get the character of an inverse 
power law.

\subsection{DEA at work: non-coding and coding DNA sequences}
By using the DEA  algorithm we can detect the existence of
scaling, either normal or anomalous, Gaussian or L\'{e}vy,  in a
very efficient way, and without altering the data with any form of
detrending. We analyze the data of both the coding and non-coding
sequences. Starting from the sequence $\xi_i, ~i=1,...,N$ we
create the diffusion trajectories and we compute the diffusion
entropy $S_{d}(l)$ according to equation (\ref{entropy}). The results
are reported in Figs. 7-9. We determine the scaling as the slope of 
the tangent of the curve $S_{d}(\tau)$.
As for the second moment scaling, called $H$ or $H'$, according to 
whether it refers to long or short times, we adopt for
the DEA scaling the corresponding symbols $\delta$ and $\delta'$. It 
is evident that $\delta$ is the true scaling. As to the meaning of
$\delta'$, it will be discussed at the end of this section.

{\bf Non-coding DNA}. 
First of all let us consider the {\it non-coding
sequence} HUMTCRADCV. Fig. 7a shows that the DEA results
in what seems to be a time dependent  scaling.  This is pointed out 
by means of the two straight lines of different slopes, 
$\delta'=0.615 \pm 0.01$ and $\delta=0.565 \pm 0.01$
Anomalous diffusion shows up at
both the short-time and the long-time scale, and this seems to be
a common characteristic of non-coding sequences, supported also
by the application of our technique to other non-coding DNA
sequences. Moreover, we notice that the scaling in the short-time
regime $\delta'=0.615 \pm 0.01$ coincides  with the value
found by means of the DFA analysis \cite{stanley3}, $H'=0.61\pm0.01$.
The authors of Ref. \cite{stanley3} assign this scaling value to
both the short and the long-time regime, while the DEA detects a 
different scaling at long times.
Fig. 7b shows the result of the DEA   applied to an
artificial sequence built up  according to the CMM prescription so
as to mimic the sequence HUMTCRADCV. 
We use both $\mu$ and $p_{R}$ as fitting parameters. 
In this case, the intensity
of the random component is not predominant,  as in the case of the
coding sequences, which are known \cite{buiattino} to require the
condition of Eq.(\ref{xxx}). In fact, in this case the best
fit between the real and the CMM sequence is obtained by
setting $p_{R} = 0.56\pm0.02$. As to $\mu$, the value of it emerging 
from this fitting procedure,  is considered by us to be the best
estimate of this inverse power law index. This value is $\mu = 
2.77\pm0.02$. If we plug it into Eq.(\ref{hursvaluedls2}),
we  get $H = 0.615\pm0.01$, which is in fact the scaling detected in 
Ref.\cite{stanley3}. This means that the short-time region obeys the 
FBM statistics. If we plug it into Eq.(\ref{levywalkdelta})
we obtain $\delta = 0.565\pm0.01$, which is the slope of the DEA 
curve in the long-time
regime, thereby proving that the relation between $\delta$, the true 
scaling, and $H$ obeys the condition of Eq.(\ref{LCcond}), which is, 
as erlier pointed out, a clear indication of L\'{e}vy statistics.
We consider this to be  a compelling evidence that at this long-time 
scale L\'{e}vy rather that FBM diffusion is generated.

{\bf Coding DNA}.
In Figs. 8 and 9 we turn to the more delicate problem of 
{\it coding sequences}. The first sequence (ECO110K) has already
been studied by means the DFA analysis in Ref. \cite{stanley3}.
The DFA finds $H' = 0.52\pm0.01$ at  the short-time scale and $H
= 0.75\pm0.01$ in the large-time scale. The second sequence
(ECOTSF) has been analyzed in Ref. \cite{barbi} by using four
different methods. The first was the SDA discussed in Section 
 C.
This is a method of analysis less sophisticated
than the DFA, since does not imply any local detrending. The
second and third methods were the DFA and the Hurst analysis
\cite{feders}, respectively. The fourth method used was the
Onsager regression analysis, a method that, in that context,
provides information on the correlation function of the
fluctuation $\xi$, which has an inverse power dependence on time
$l$ with the power index $\beta = \mu-2$. The authors of Ref.
\cite{barbi}, by using essentially the first method and the
Onsager regression analysis, reached the conclusion that the most
plausible value of the scaling parameter in the long-time region
is $H = 0.75\pm0.01$ that is equivalent to the exponent
$H=0.74\pm0.01$ found in Figs. 6. It is interesting to remark
that the coincidence among the different predictions about
scaling, and especially that between the second moment technique
and the Hurst analysis,  implies the adoption of the Gaussian
assumption \cite{mannella}. On the other hand, when that condition
does not apply and the two scaling predictions are different, to
the best of our knowledge, it does not seem to be known what is
the meaning of any of them. Furthermore, the authors of Ref.
\cite{buiattino} pointed out that the statistics of the long-time
regime is too poor to support any claim on the departure from the
Gaussian condition. In conclusion, in Ref. \cite{buiattino} the claim 
that the DNA statistics is
of L\'{e}vy kind was  essentially based on the assumption that the 
dynamical theory of Refs.\cite{barbi,allegro} is a reliable approach 
to the statitstics of DNA sequences.
No direct evidence was provided.

The DEA  method allows us to prove that the conjecture of the authors 
of Ref. \cite{buiattino} is correct: the results illustrated in Figs. 
8 and 9 afford a
convincing proof that the DNA statistics is
of L\'{e}vy kind}. Figs. 8a and 9a clearly show the difference
between the slope at short time 
(respectively $\delta'=0.52\pm0.01$ in Fig. 8a and 
$\delta'=0.53\pm0.01$ in Fig. 9a) which, in this case, is very
close to that of ordinary random walk, and the slope at long time
that corresponds to $\delta= 0.665\pm0.01$. Since we know that in
both cases the  long-time slope provided by the DFA is $H =
0.75\pm0.01$, we conclude that in both cases the condition of Eq.
(\ref{LCcond}), indicating L\'{e}vy statistics,  applies.
Figs. 8b and 9b aim at fitting the  curves produced by the DEA 
method, applied to the real sequences by means of the CMM model.
The purpose is not only that of proving that the CMM can become so 
close to the real results as to be virtually indistinguishable from 
them. It is also a way, already applied in Fig.7b,  to derive very 
accurate values for the power index $\mu$.   A very good agreement is 
obtained by
setting $p_{R} = 0.943\pm0.01$ for ECO110K (Fig.5b) and $p_{R} = 0.937\pm0.01$
for ECOTSF (Fig.7b). The very good fitting accuracy supports the physical
reasons that led the authors of Ref. \cite{barbi} to propose the
CMM model for coding sequences. In fact, with the  large weight, 
$p_{R} = 0.937\pm0.01$,
assigned to the random component,  the scaling values become
$\delta'=0.52\pm0.01$ and $\delta'=0.53\pm0.01$, namely, very close to the
conventional scaling $\delta=H = 0.5$. This normal condition  lasts 
for an extended
period of time, and eventually,  at larger times the transition to a 
larger scaling
takes place.

We note that the authors of Ref. \cite{arneodo} find anomalous
diffusion in a statistical condition that they claim to be
Gaussian.
  According to the result of Ref.\cite{allegro},
the Gaussian condition is incompatible with a
stationary diffusion process generated by a dichotomous fluctuation
yielding a non integrable correlation function with an inverse power
law character. This dichotomous fluctuation is expected to generate
L\'{e}vy rather than Gauss statistics. 
The authors of Ref. \cite{paolophdthesis} studied under 
which physical condition
FBM is allowed to show up, in apparent conflict with 
the conclusions of Ref. \cite{allegro}. 
With the 
help of a fractal model for the DNA folding, the authors of Ref. \cite{paolophdthesis} proved that  FBM, 
advocated by the paper of Ref. \cite{arneodo}, is possible as
a form of non-stationary process. Thus, in principle, the arguments 
of the work of Ref.\cite{allegro} would not rule out the 
possibility that
the changing slope is a manifestation of a FBM 
with a time dependent scaling. This would be another
form of transition from dynamics to thermodynamics, 
of extremely large time duration.
However, this way of establishing a compromise between the compelling 
prediction of the GCLT, according to which a dichotomous process 
with long-range correlation ($2 < \mu <2$ must produce L\'{e}vy 
statistics, and the conclusion of some authors that this statistics 
is Gaussian\cite{arneodo} is ruled out by  the statistical analysis 
of the present paper, which is made
much more accurate than the earlier approaches by the DEA.
This is made compelling not only by the results illustrated in
Figs. 8 and 9, but also by a plenty of statistical measurements on different 
DNA sequences, reported in Table II.
All these results prove that the equality of Eq.(\ref{LCcond}),
implying L\'{e}vy statistics, applies to both kind of
sequences. This means that in both cases the long-time limit is
characterized by L\'{e}vy  statistics and that this is the form of
non-Gaussian statistics revealed by the analysis of Ref.\cite{buiattino}.

We can now address the delicate problem of the transition from 
$\delta'$ to $\delta$. On the basis of the  results of Figs. 8 and 9, 
we would be tempted to conclude hat the CMM is a
reliable dynamical model for DNA sequences. If this is correct, the 
transition from $\delta'$ to $\delta$ is really a time dependent 
scaling. In fact, according to the CMM the
short-time scale is dominated by the random component, due to the 
fact that the condition of Eq.(\ref{xxx}) applies. In the case of 
Fig. 7 the transition from $\delta'$ to $\delta$ is probably 
dominated by a completely different effect. This is the slow 
transition from dynamics to thermodynamics discussed in
Section \ref{sectlevy}

\subsection{Significance of the results obtained}

To properly appreciate the significance of the results of this section,
    it is necessary to say a few words about the two different scaling
    prescriptions of Eqs.(\ref{levywalkdelta}) and 
(\ref{hursvaluedls2}). The scaling prescription of
    Eq.(\ref{hursvaluedls2}) is determined by the adoption of the 
variance method,
    as clearly illustrated by the dynamical approach to the DNA sequences
    of Ref. \cite{barbi}. 
This prescription is not ambiguous if the condition of
    Gaussian statistics applies. In fact, a Gaussian distribution drops
    quickly to zero, and the existence of a finite propagation front does
    not produce any significant effect. It has to be pointed out, in fact,
    that the adoption of the Brownian landscape proposed in the pioneer
    papers of Refs. \cite{stanley1,stanley2,stanley3}
    implies the existence of a
    propagation front moving with ballistic scaling ($\delta = 1$). In
    other words, if we find a window of length $l$ filled with only $1$ 's
    or with only $-1$'s, this means a trajectory travelling with uniform
    velocity, and the x-space at distances from the origin larger that $l$
    is empty. The existence of a propagation front does not have big
    consequences in the case of Gaussian statistics, since the population
    at the propagation front is essentially zero in that case. It is not so
    in the case of L\'{e}vy statistics, though, due to the existence of very
    long tails in that case. Therefore the L\'{e}vy processes resulting from
    these sequences are essentially characterized by the presence of two
    distinct scaling prescriptions, the L\'{e}vy prescription of Eq.
(\ref{levywalkdelta}),
    concerning the portion of distribution enclosed between the two
    propagation fronts, and the scaling $\delta = 1$, of the propagation
    front itself. The scaling of the variance of Eq.
(\ref{hursvaluedls2}) does not
    reflect correctly either of these two different scaling prescriptions,
    being a kind of compromise between the two. The scaling of the
    distribution enclosed by the two propagation fronts is, on the
    contrary, a genuine property that corresponds to the prediction of the
    GCLT\cite{gnedenkokolmogorov}. It is very
    satisfactory indeed that the DEA method makes this genuine form of
    scaling emerge. Furthermore, the DEA is a very accurate method of
    scaling detection, as proved by the fact that it reveals the existence
    of L\'{e}vy statistics in the case of the coding sequence. In this case,
    as pointed out by the authors of Ref. \cite{barbi}, the ordinary methods
    become inaccurate due to the poor statistics available in the long-time
    limit.

    Another important result of this section is that it
    confirms the validity of the CMM model. This model is expected to
    generate L\'{e}vy statistics not only in the case of non-coding sequences,
    where it is easier to reveal this property. It predicts L\'{e}vy statistics
    also in the case of coding sequences as the one here analyzed.
In Ref. \cite{barbi} the emergence of L\'{e}vy statistics was
conjectured but not proved, due to the fact that in that paper the
observation was made monitoring the probability distribution
$p(x,t)$. As already pointed out, the lack of sufficient statistics 
makes it difficult  to assess if the distribution $p(x,t)$ has, or 
not,  tails with an inverse power law character.
  In Ref. \cite{buiattino} a clear deviation
from the Gaussian condition was detected in the long-time limit,
but, again, no direct evidence was found that this deviation from
Gaussian statistics takes the form of L\'{e}vy
statistics. The results of this section  prove, with the help of
the artificial sequences of Section \ref{secttrans}, that the DEA  is
a method of analysis  so accurate as to assess with good accuracy the
property of Eq.(\ref{LCcond}), and with it, the emergence of 
L\'{e}vy statistics for both coding and non-coding sequences.

In conclusion, this paper lends support, with the help of the DEA, 
an efficient technique of scaling detection, to the claims 
of Allegrini \emph{et al} \cite{allegro} about the controversy 
between Voss\cite{voss} and the authors
of Ref.\cite{stanley1}. 
The differences in the findings of the groups-long-range 
correlations being ubiquitous in DNA sequences by Voss\cite{voss} 
and such correlations being absent in Ref.\cite{stanley1},
motivated the authors of Ref. \cite{allegro} to develop a 
phenomenological model, the CMM, that might have mitigated 
the differences between the two apparently conflicting perspectives. 
The validity of the point of view of these authors is fully confirmed, 
since L\'{e}vy statistics, and consequently long-range correlations, 
seem to be ubiquitous, being a property of the long-time regime 
of both coding and non-coding sequences,
while the properties of ordinary Brownian motion are confined 
to the short-time regime of coding sequences. 

\section{CONCLUSIONS}

This paper shows that long-range correlations result 
in a very slow transition to scaling, regarded as a form of
thermodynamic equilibrium. 
The standard methods of statistical analysis (variance methods) 
are a source of misleading information in this case: the first  
being mistaking the regime of transition 
to scaling as either ordinary or anomalous scaling.
The second is that the scaling value, as determined by the evaluation 
of the second moment, might significantly depart from the 
correct one.  
All the VM techniques are shown to be affected by this limitation,
while the DEA is the only technique always yielding the correct scaling 
value, if the scaling condition applies. 
The application to the study of DNA sequences reported 
in this paper yields: 

1) a striking example 
of how the standard techniques can produce
misleading conclusions 

2) a suggestive example of the power of the DEA, which,  
in this case,  is able to indicate clearly that both coding 
and non-coding sequences generate L\'{e}vy statistics 
in the long-time limit. 

The DEA is not only a method of scaling detection. Its entropic nature 
gives also useful insights into the regime of transition 
from dynamics to thermodynamics. 
It is possible to prove that in the special case
where the time series is generated 
by fluctuations around a locally varying bias
the regime of transition to the scaling regime is significantly delayed.
The ordinary techniques of analysis mistaken this transition 
regime as a form of anomalous memory, while
the DEA makes it possible to establish the genuine nature 
of the process under study. 
This is left as a subject for further applications.


\onecolumn

\newpage
\begin{figure}[h]  
\epsfig{file=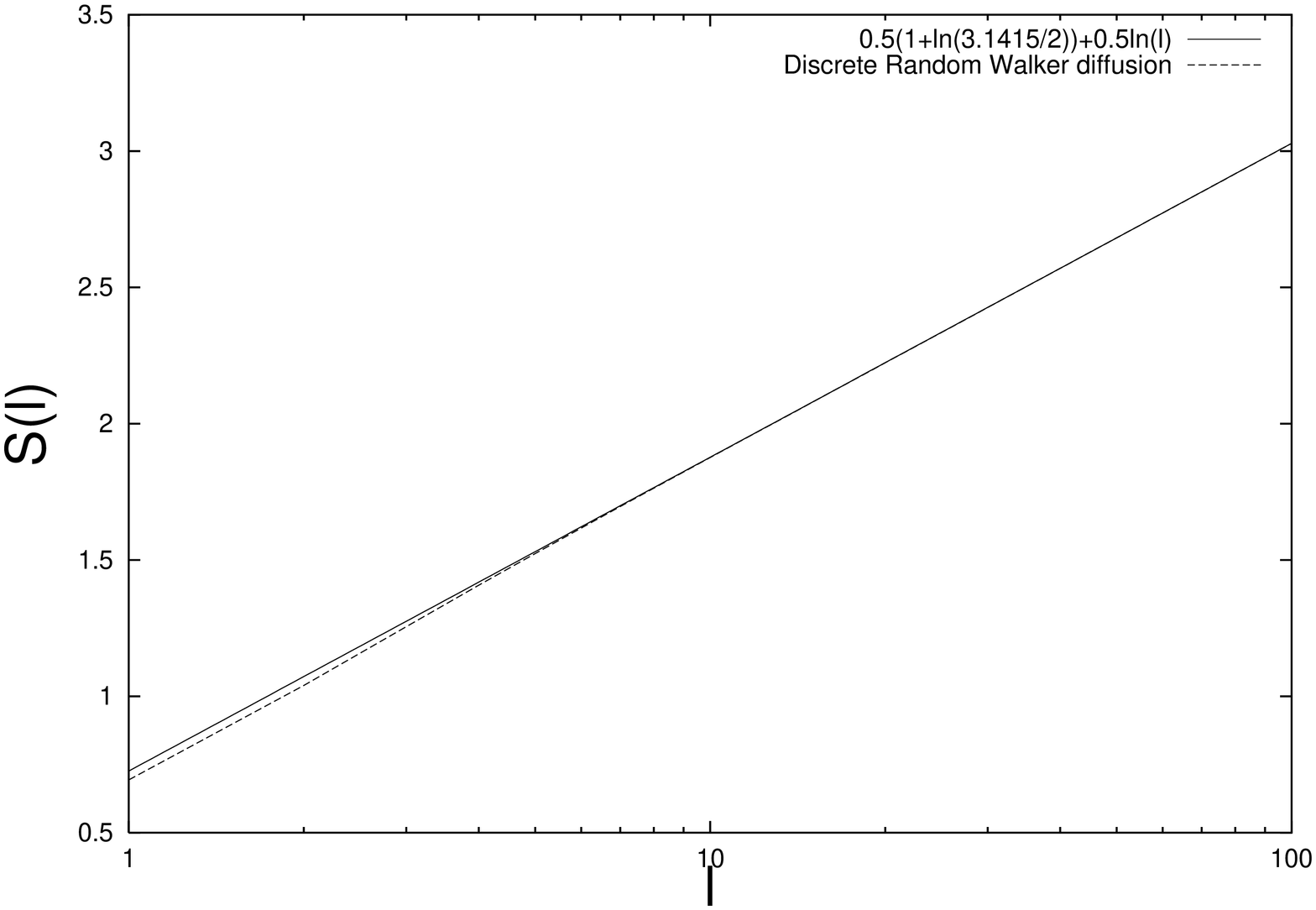, height=16cm,width=12cm,angle=-90}
\caption{
Diffusion entropy of a random walker as a function of the number of jumps
l. The dashed lines and the solid line denote the discrete diffusion
entropy $S_{d}(l)$, of Eq.(\ref{exactinthediscrete})
and the continuous prescription of Eq.(\ref{continuousshannonentropy}), respectively. After a short transient the dashed line converges to the solid line. 
}
\end{figure}

\newpage
\begin{figure}[h]  
\epsfig{file=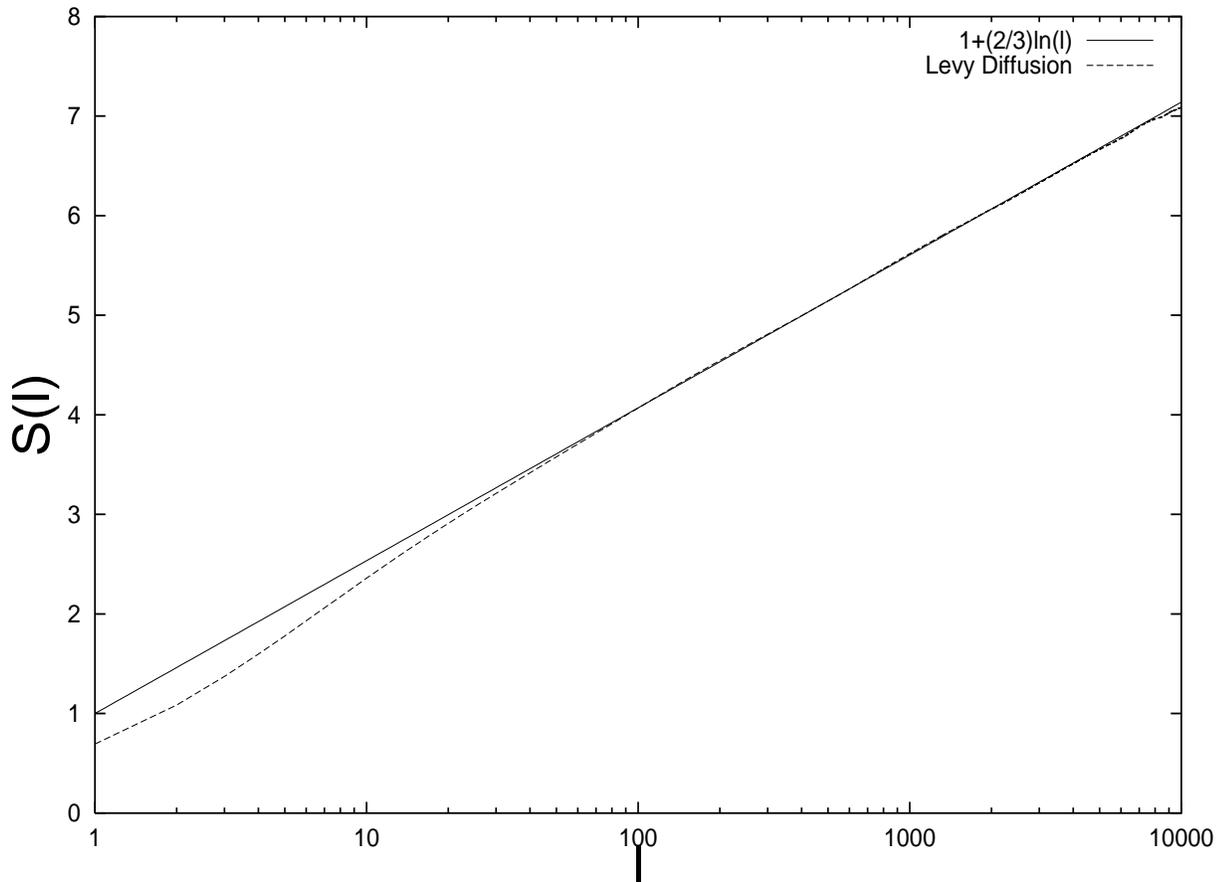,height=16cm,width=12cm,angle=-90}
\caption{Diffusion entropy of the L\'{e}vy process generated by an
articial sequence, $ \xi _i$, corresponding to the power coefficient
$\mu=2.5$ and $T=0$. The dashed line is the diffusion entropy, $S_d(l)$,
in the discrete-space perspective given by the  Eq.(\ref{entropy}).
The solid line is the diffusion entropy  $S(l)$
in the continuous-space perspective given by the
Eq.(\ref{Levyfitting}). After an initial 
transient, the dashed line converges to the solid line.  
}
\end{figure}

\newpage
  \begin{figure} \label{fig59de}   
\epsfig{file=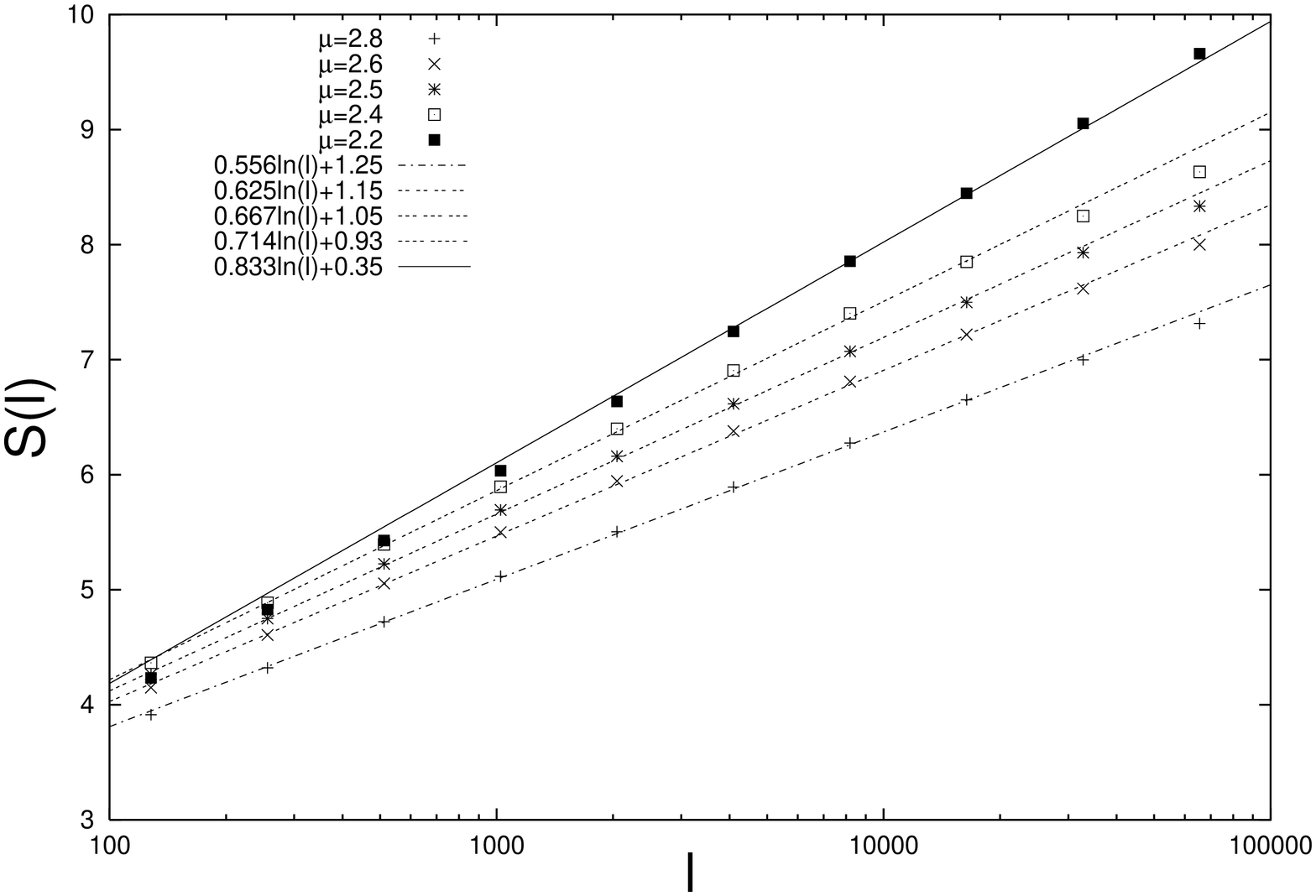,height=16cm,width=10cm,angle=-90}
\epsfig{file=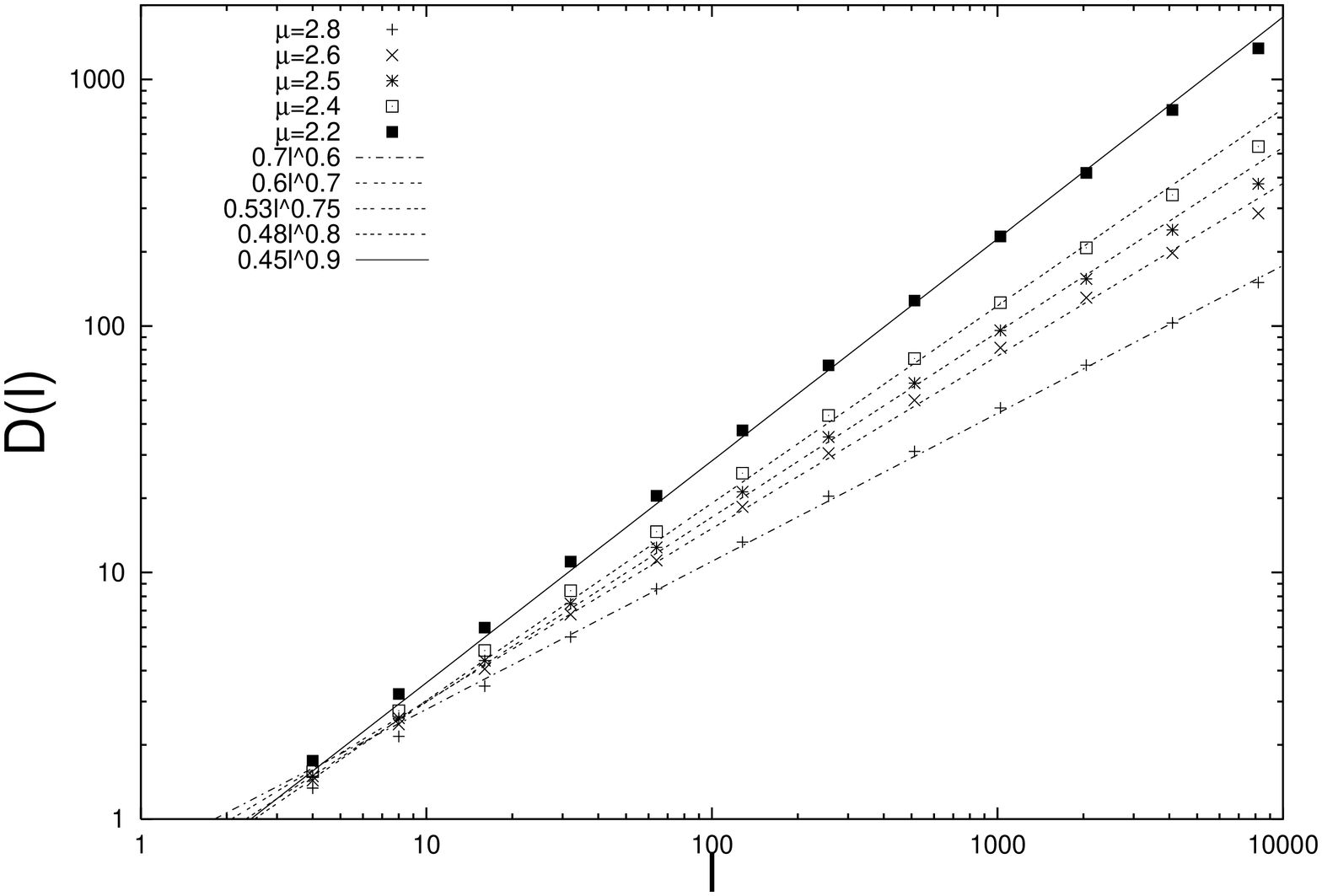,height=16cm,width=10cm,angle=-90}
\caption{ Diffusion entropy and Standard Deviation 
of the L\'{e}vy process generated by articial sequences
$\xi _i$ corresponding respectively to five
different values of the power coefficient $\mu$, namely:
$\mu=2.8, 2.6, 2.5, 2.4, 2.2$, and $T=0$.
The numerical results of the Diffusion Entropy Analysis (DEA) (3a)
and of the Standard Deviation Analysis (SDA) (3b), reported in symbols,
are in perfect agreement with the theoretical predictions,
reported as fitting lines and obtained by using
respectively the values of the pdf scaling
exponent $\delta$ and of the esponents $H$ in Table I.
}
\end{figure}

\newpage
\begin{figure}     
\epsfig{file=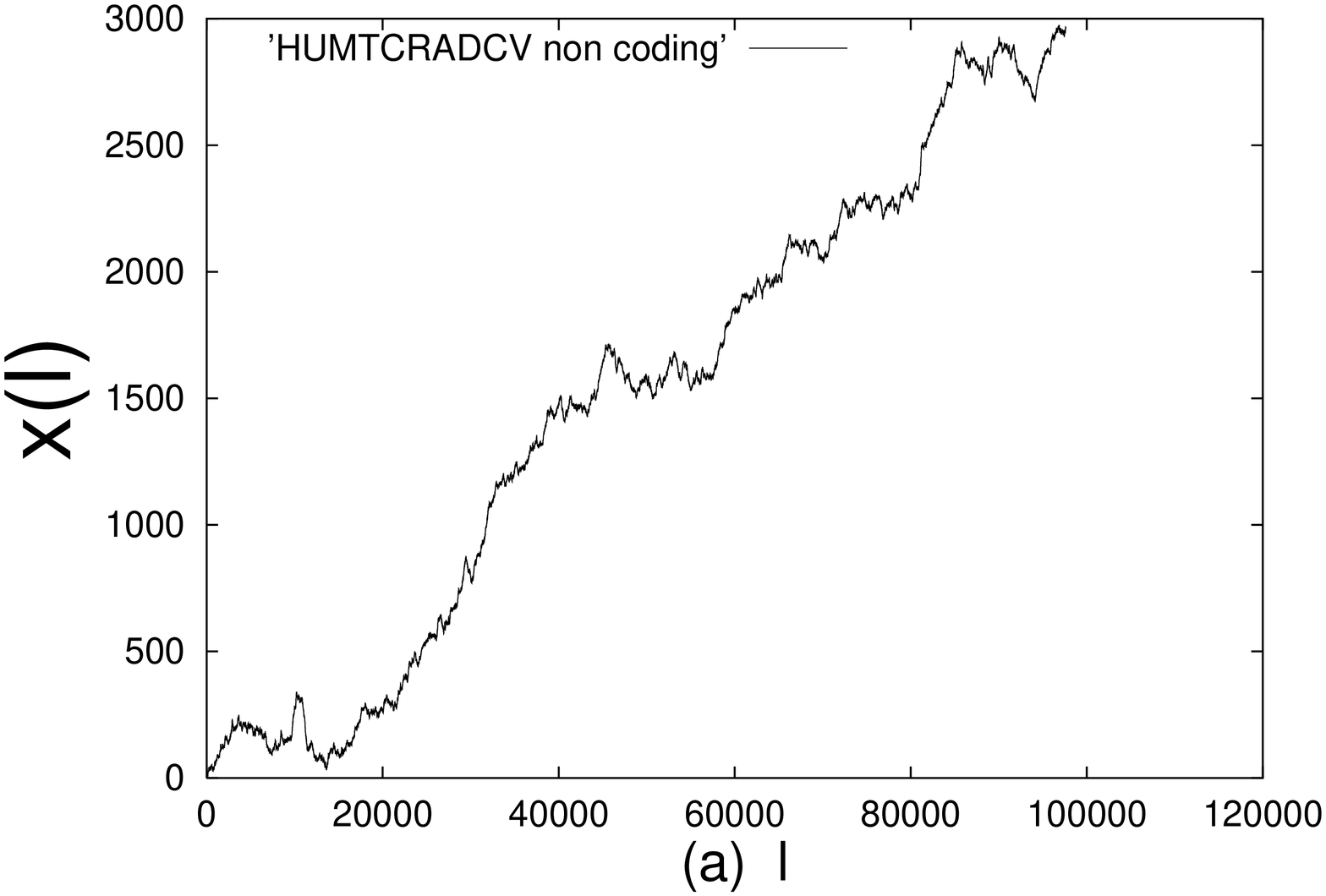,height=14cm,width=5cm,angle=-90}\\
\epsfig{file=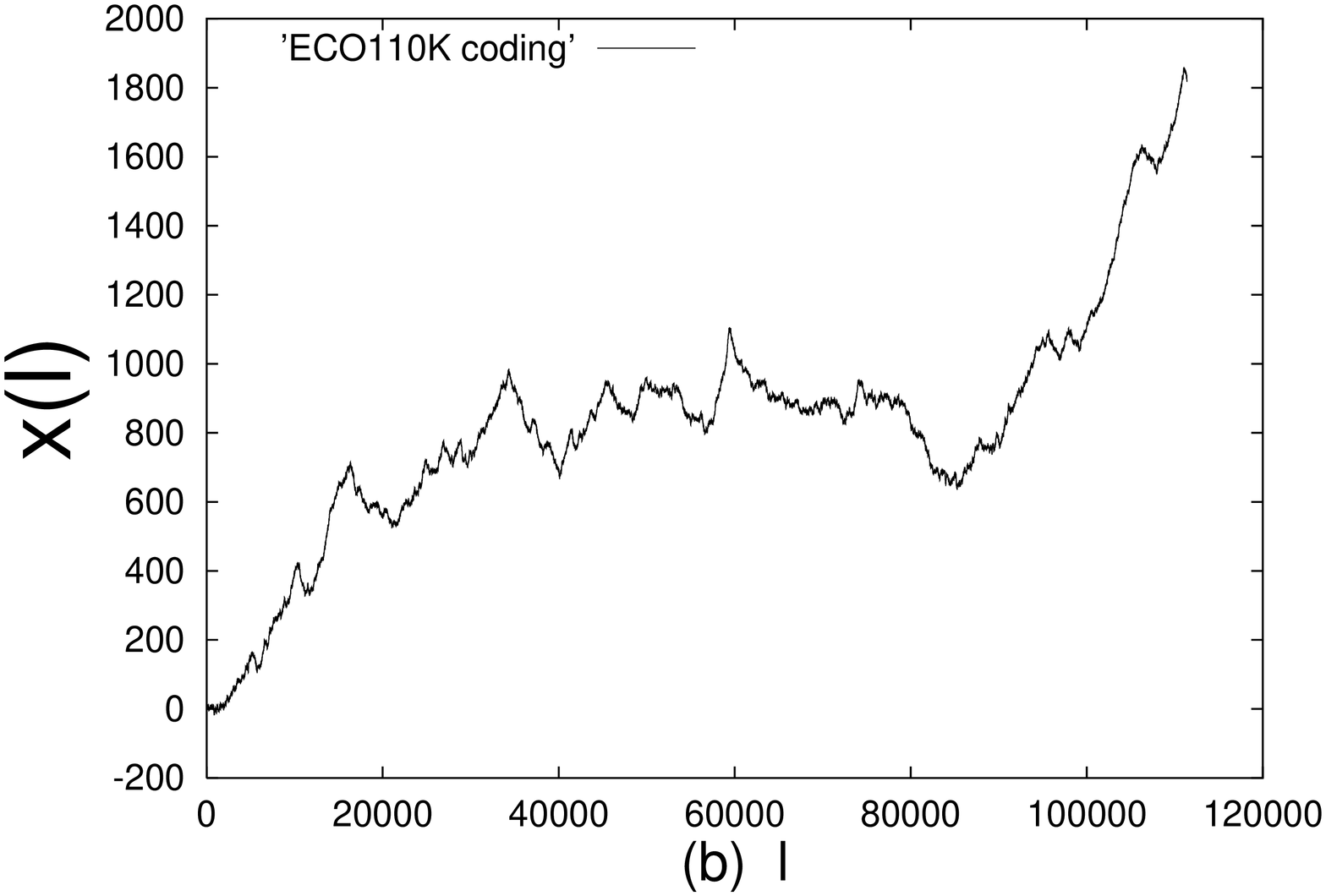,height=14cm,width=5cm,angle=-90}\\
\epsfig{file=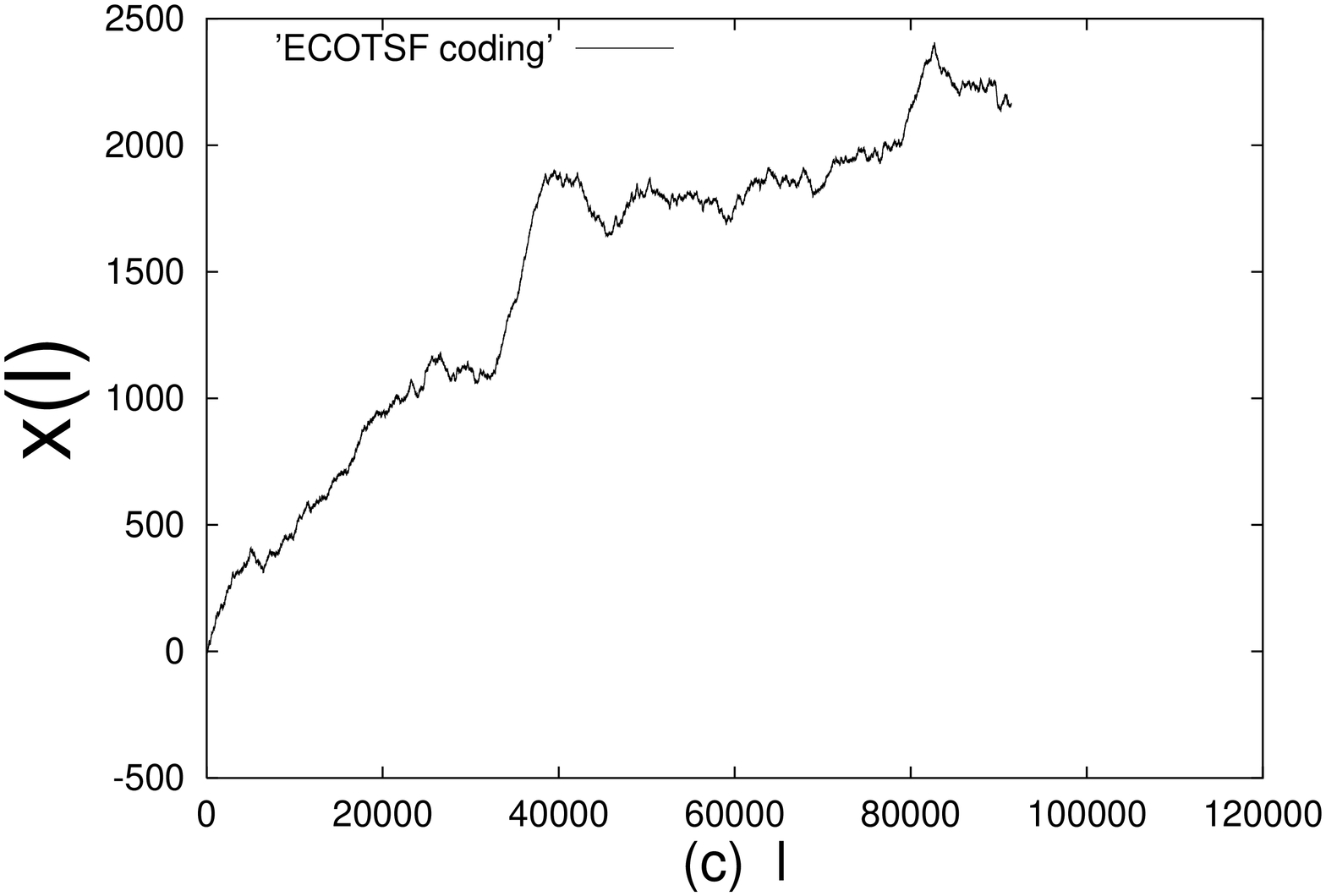,height=14cm,width=5cm,angle=-90}
\caption{ In (a) we report the DNA walk relative to
HUMTCRADVC, a non-coding chromosomal fragment. In (b) and (c),
we report the DNA walk relative to ECO110K and ECOTSF,
two coding genomic fragments. }
\end{figure}

\newpage
\begin{figure}   
\epsfig{file=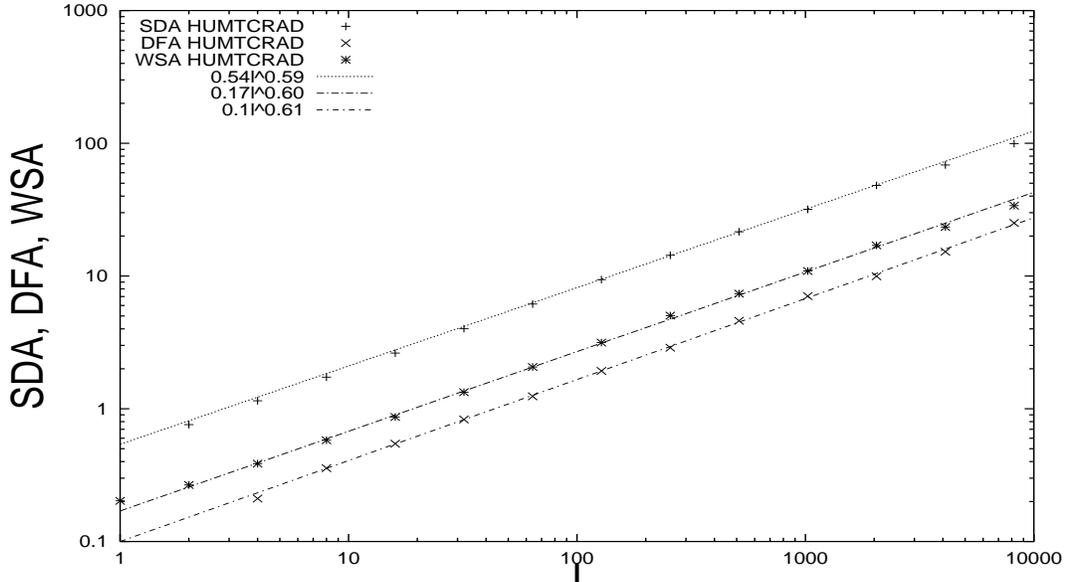,height=14cm,width=8cm,angle=-90}
\caption{Application of the three variance methods SDA, DFA and
WSA to the sequence HUMTCRADVC, the
non-coding chromosomal fragment. The three methods give the same 
exponent $H$. In fact we get $H=0.59\pm0.01$ (SDA), $H=0.60\pm0.01$ 
(DFA) and $H=0.61\pm0.01$ (WSA), where the differences are within the error bars. 
Moreover $H$ is the same both at short-time and long-time regions (i.e. $H'=H$). 
}
\end{figure}

\newpage
\begin{figure}   
\epsfig{file=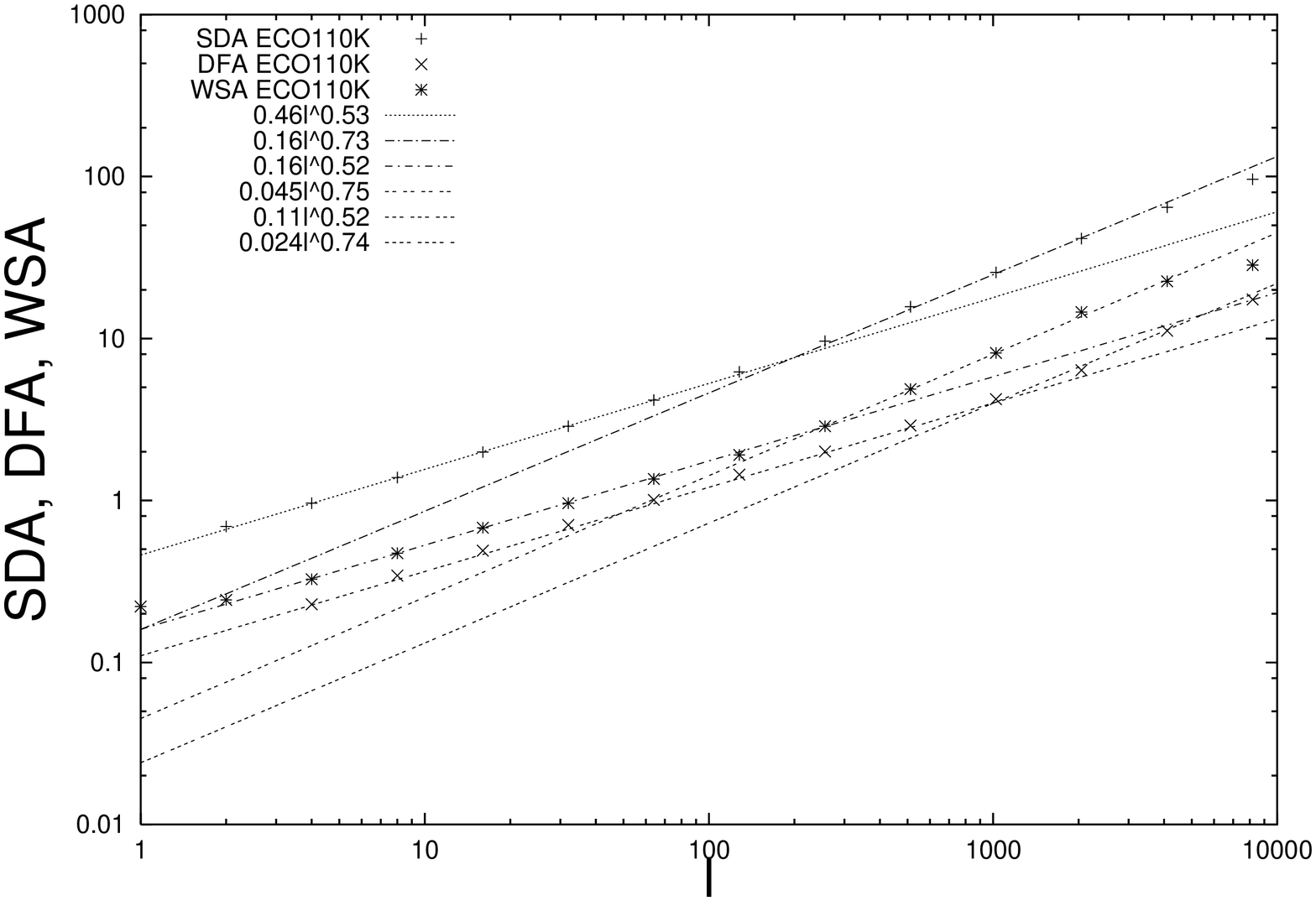,height=14cm,width=8cm,angle=-90}
\epsfig{file=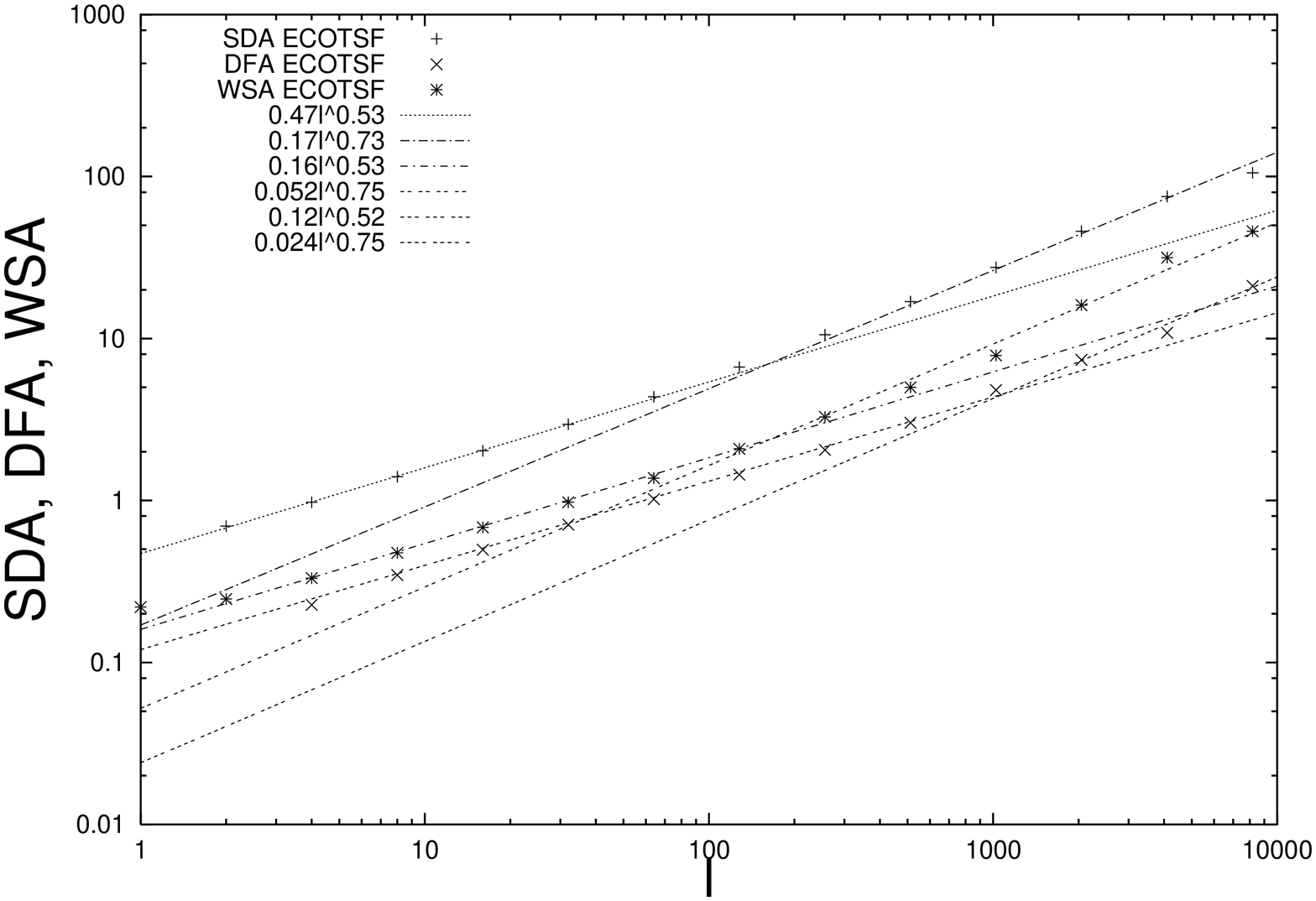,height=14cm,width=8cm,angle=-90}
\caption{
Application of the three variance methods
SDA,  DFA and WSA to ECO110K (a) and ECOTSF (b), the two coding
genomic fragments.
The scaling exponent $H'$ in the short-time region is $0.53\pm0.01$ 
(SDSA), $0.52\pm0.01$ (DFA), $0.52\pm0.01$ (WSA).
The scaling exponent $H$ in the long-time regions is $0.73\pm0.01$ 
(SDSA), $0.75\pm0.01$ (DFA), $0.74\pm0.01$ (WSA).
}
\end{figure}

\newpage
\begin{figure}   
\epsfig{file=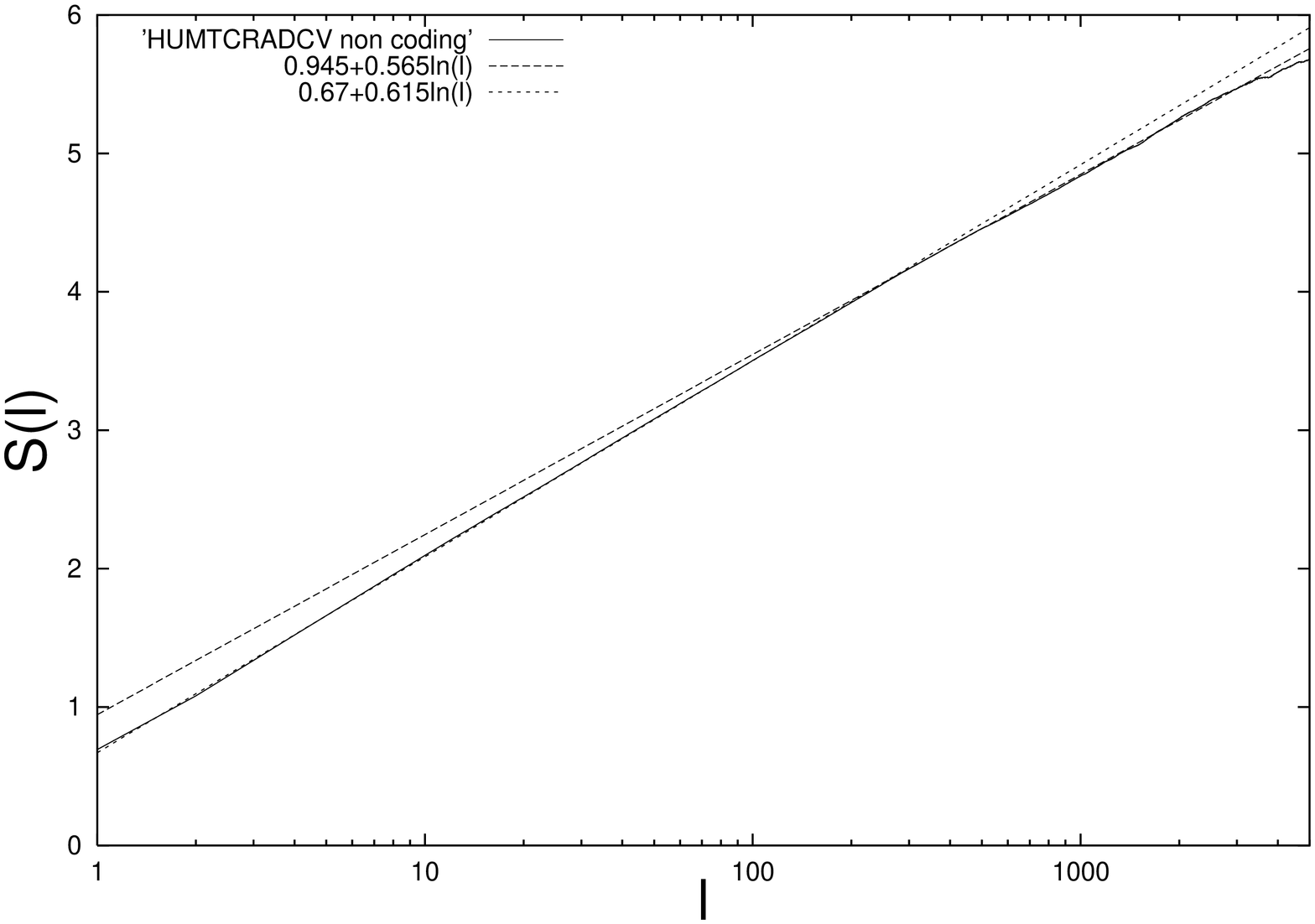,height=14cm,width=8cm,angle=-90}
\epsfig{file=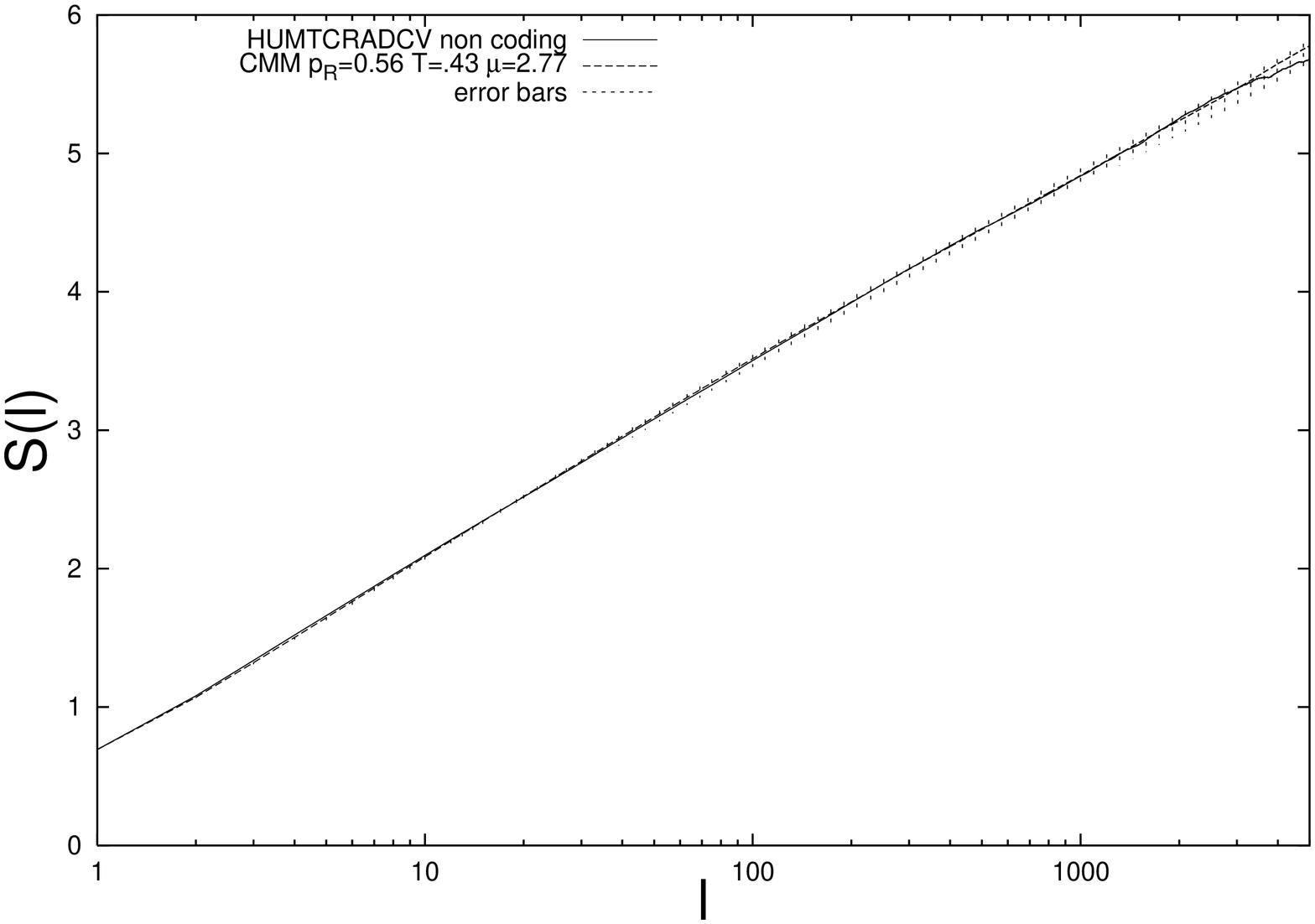,height=14cm,width=8cm,angle=-90} \caption{
Diffusion Entropy for the HUMTCRADCV (the
non-coding chromosomal fragment) and its CMM simulation.
Fig.7a shows that the DE
analysis results in a scaling changing with time. The slope of the
two straight lines is $\delta'=0.615\pm0.01$ in the short-time region,
and $\delta=0.565\pm0.01$ in the long-time regime.
Fig.7b shows the comparison between the DEA of the real non-coding
sequence and an artificial sequence corresponding to the CMM
model: $p_R=0.56\pm0.02$, $T=0.43$, $\mu=2.77\pm0.02$. }
\end{figure}

\newpage
\begin{figure}  
\epsfig{file=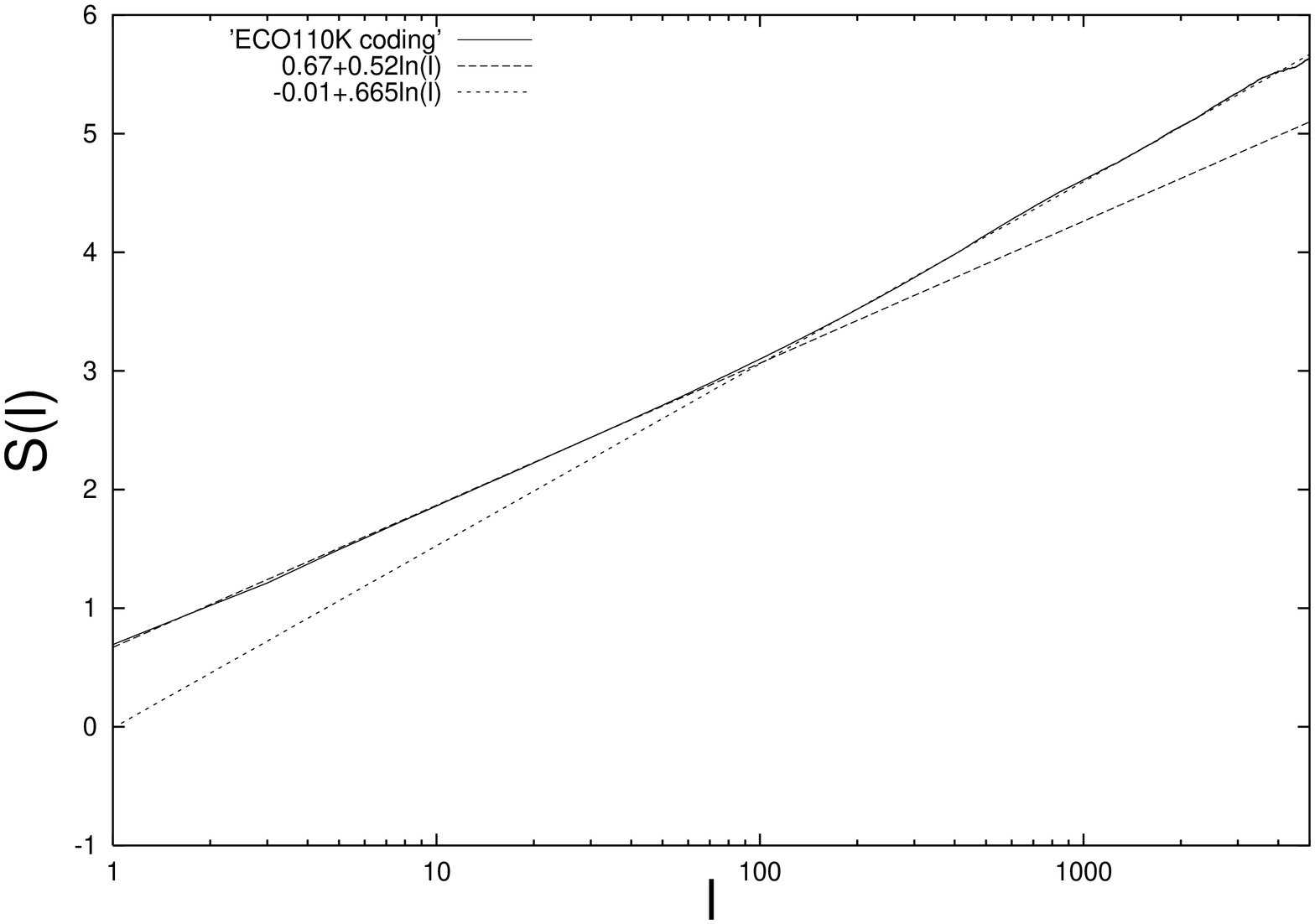,height=14cm,width=8cm,angle=-90}
\epsfig{file=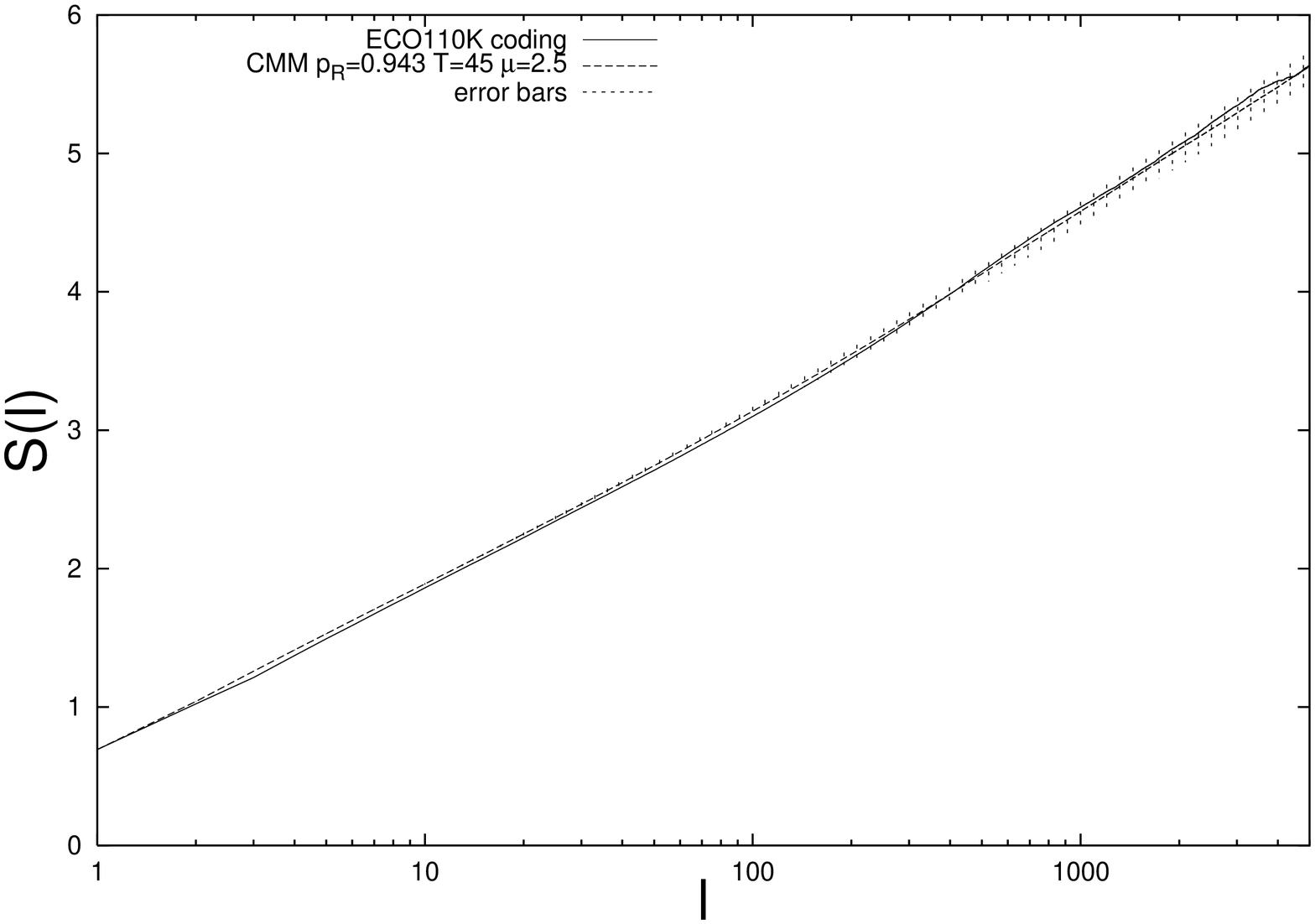,height=14cm,width=8cm,angle=-90}
\caption{
Diffusion Entropy for the ECO110K (one of the two coding
genomic fragments studied) and its CMM simulation.
Fig.8a shows that the DEA  results in
a scaling changing with time.
The slope of the two straight lines
is $\delta'=0.52\pm0.01$ at short-time regime,
and $\delta=0.665\pm0.01$
at long-time regime. Fig.8b shows the comparison between the DE
analysis of the real  coding sequence and an artificial sequence
corresponding to the CMM model: $p_R=0.943\pm0.01$, $T=45$, 
$\mu=2.5\pm0.02$.  }
\end{figure}

\newpage
\begin{figure}    
\epsfig{file=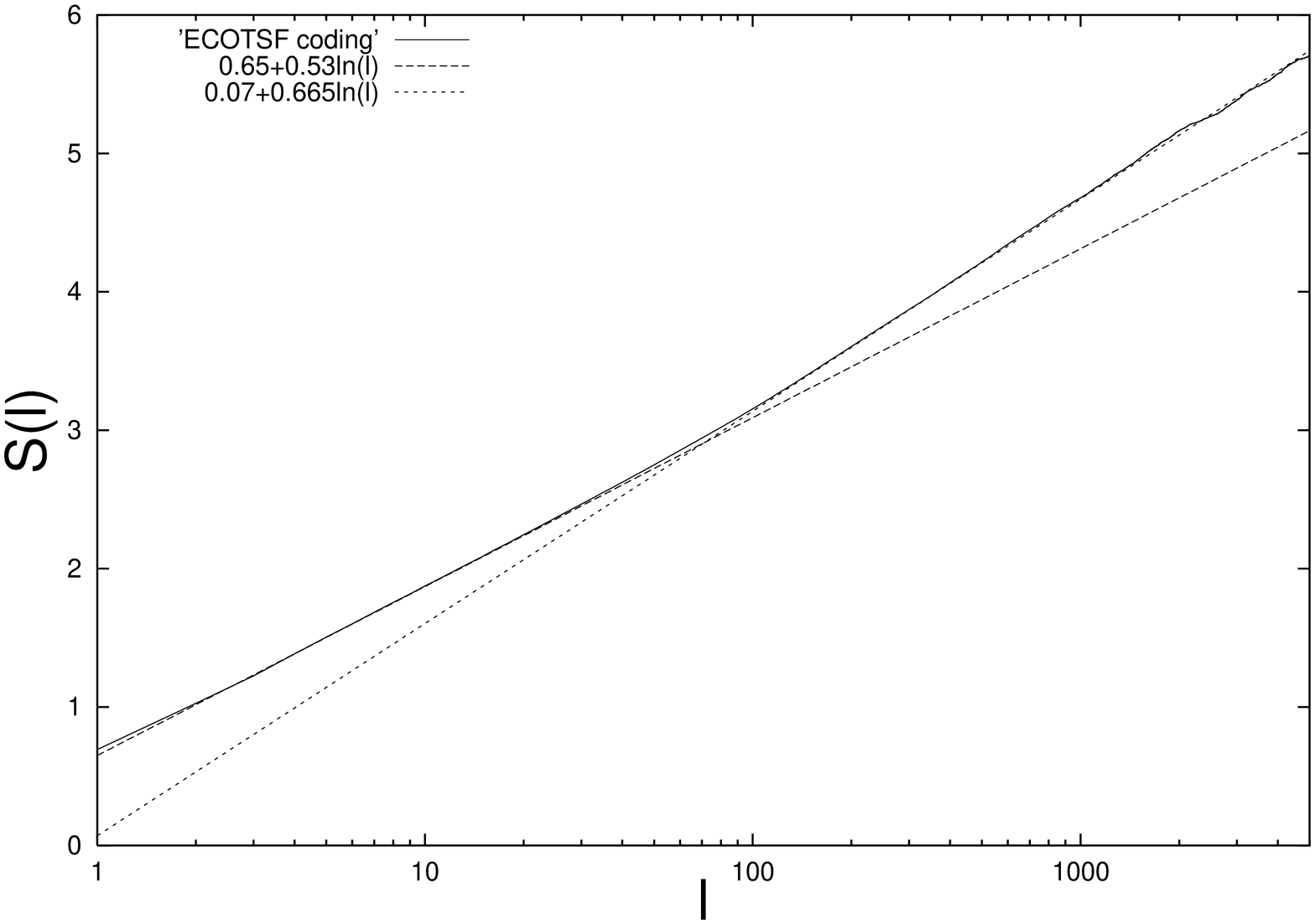,height=14cm,width=8cm,angle=-90}
\epsfig{file=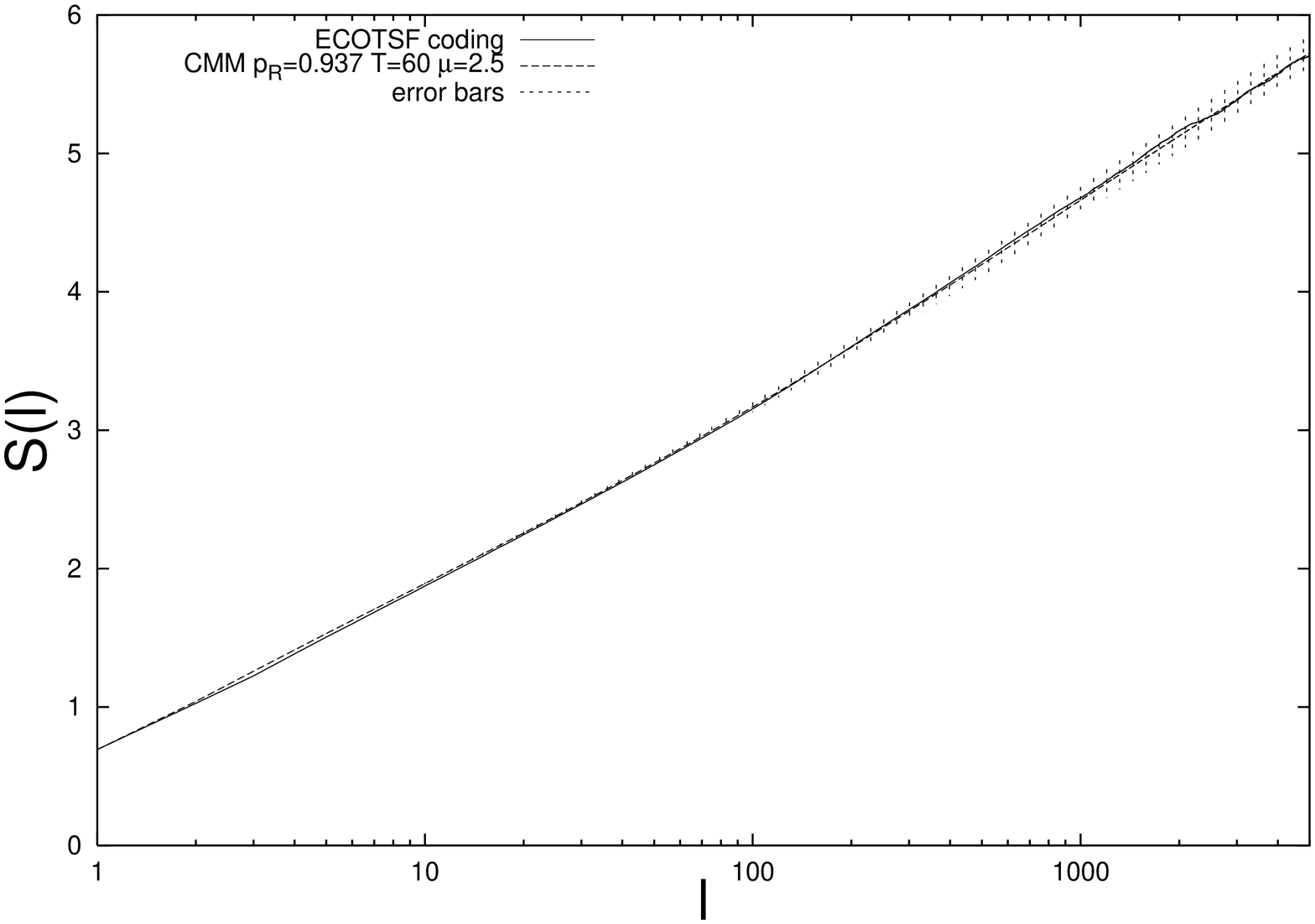,height=14cm,width=8cm,angle=-90}
\caption{
Diffusion Entropy for the ECOTSF (the second of the two coding
genomic fragments studied in this paper) and its CMM simulation.
Fig.9a shows that the DEA results in
a scaling changing with time. The slope of the two straight lines
is $\delta'=0.53\pm0.01$ in the short-time region, and
$\delta=0.665\pm0.01$ in the long-time regime. Fig.9b shows the
comparison between the DEA   of the real  coding sequence
and an artificial sequence corresponding to the CMM model:
$p_R=0.937\pm0.01$, $T=60$, $\mu=2.5\pm0.02$.  }
\end{figure}

\newpage
\begin{table}    
\begin{center}
   \begin{tabular}{|c|c|c|c|c|c|}\hline
  $\mu$         & 2.200     &  2.400    &  2.500    & 2.600     & 
2.800     \\ \hline
  $H$         &  0.900    & 0.800     & 0.750     &  0.700    &  0.600 
\\ \hline
  $\delta$         &  0.833    &  0.714    &  0.667    &  0.625    & 
0.556    \\ \hline
   \end{tabular}
\end{center}
\caption{In the first line from the top we report the power indices 
of the inverse power law distributions used to create 
the artificial sequences studied in Fig.3. 
In the second line from the top we report the
corresponding Hurst coefficients, 
prediction of Eq.(\ref{hursvaluedls2}). 
In the third line from the top we report the true scaling, 
namely the L\'{e}vy scaling of Eq.(\ref{levywalkdelta}).}
\end{table}

\newpage
\begin{table}   
\begin{tabular}{|c|c|c|c|c|}\hline
Non-Coding            & N       &  $H$ &$\delta_{H}$     &  $\delta$  \\ \hline
HUMTCRADCV  & 97630   &  0.61   &  0.56  &      0.56\\
CELMYUNC & 9000 & 0.71 & 0.63 &0.635\\
CHKMYHE  & 31109   &  0.78   &  0.69  &     0.70\\
DROMHC  & 22663   &  0.72   &  0.64  &      0.65\\
HUMBMYHZ  & 28437   &  0.58   &  0.54 &     0.54\\   \hline
Coding          &        &  &  &   \\ \hline
  ECO110K    &  111401 &   0.74     &  0.66&     0.66\\
  ECOTSF     &  91430  &       0.74     &  0.66&     0.66\\
LAMCG      &  48502  &   0.85     & 0.77&      0.76\\
CHKMYHN      &  7003  &   0.74     & 0.66&      0.66\\
DDIMYHC      &  6680  &   0.68     & 0.61&      0.61\\
DROMYONMA      &  6338  &   0.69     & 0.62&      0.64\\
HUMBMYH7CD      &  6008  &   0.63     & 0.57&      0.58\\
HUMDYS&  13957  &   0.69     & 0.62&      0.62\\  \hline
\end{tabular}
\caption{ Values of the scaling exponents $H$ and $\delta$ for a set
of different coding and non-coding sequences.
In the first column we report the GenBank name of the
sequence \protect\cite{genbank}, and in the second column the
length $N$ of the sequence.
For all measures the error is $\pm0.01$.
$\delta_{H}$ in the fourth column is the theoretical value for
$\delta$ if the L\'{e}vy ondition applies, Eq.(\ref{LCcond}). If
the length of the genome is larger than 20,000 the fitted region
is $100<l<2000$.  If the length of the genome is shorter than
20,000, the statistics are not very good for large $l$. In this
case, the fitted region is $20<l<200$.}
\end{table}

\end{document}